\RequirePackage{fix-cm}
\documentclass[smallextended]{svjour3}       % onecolumn (second format)
\smartqed  % flush right qed marks, e.g. at end of proof
\usepackage{graphicx}
\usepackage{natbib}
\usepackage{amsfonts}
\usepackage{amsmath}
\usepackage{amssymb}
\usepackage{bm}
\usepackage{subcaption}
\usepackage{multirow}
\usepackage{url}
\makeatletter
\let\cl@chapter\undefined
\makeatletter
\usepackage{cleveref}
\usepackage[title]{appendix}
\usepackage{siunitx}
\usepackage[shortlabels]{enumitem}
\usepackage{xcolor}
\definecolor{cincinnati-red}{RGB}{190,0,0}

\newcommand{\calX}{\mathcal{X}}
\newcommand{\vecX}{\mathbf{X}}
\newcommand{\vecY}{\mathbf{Y}}
\newcommand{\vecW}{\mathbf{W}}
\newcommand{\vecC}{\mathbf{C}}
\newcommand{\Rset}{\mathbb{R}}
\newcommand{\bpi}{\boldsymbol{\pi}}
\newcommand{\bPi}{\mathbf{\Pi}}
\newcommand{\matm}{\mathbf{M}}
\newcommand{\matsig}{\mathbf\Sigma}
\newcommand{\matPsi}{\mathbf\Psi}
\newcommand{\tr}{\,\mbox{tr}}
\def\bDelta{\boldsymbol{\Delta}}
\newcommand{\bGamma}{\mathbf{\Gamma}}
\newcommand{\bPhi}{\mathbf{\Phi}}
\newcommand{\bphi}{\boldsymbol{\phi}}
\newcommand{\bI}{\boldsymbol{I}}
\newcommand{\bF}{\boldsymbol{F}}
\newcommand{\bG}{\boldsymbol{G}}
\newcommand{\bH}{\boldsymbol{H}}

\newcommand{\bL}{\boldsymbol{L}}
\newcommand{\bone}{\boldsymbol{1}}
\newcommand{\vecTheta}{\boldsymbol{\Theta}}
\newcommand{\vectheta}{\boldsymbol{\theta}}
\newcommand{\vecOmega}{\boldsymbol{\Omega}}
\newcommand{\Sa}{\mathcal{S}}
\def\bz{\boldsymbol{z}}

\begin{document}

\title{
Parsimonious Hidden Markov Models for Matrix-Variate Longitudinal Data\\
}


\author{Salvatore D. Tomarchio  \\
        Antonio Punzo    \\
				Antonello Maruotti %etc.
}


\institute{Salvatore D. Tomarchio \at
           Dipartimento di Economia e Impresa, Università degli Studi di Catania, Catania, Italia \\
           \email{daniele.tomarchio@unict.it}
           \and
           Antonio Punzo \at
           Dipartimento di Economia e Impresa, Università degli Studi di Catania, Catania, Italia \\	
					 \and
           Antonello Maruotti \at
           Dipartimento di Giurisprudenza, Economia, Politica e Lingue Moderne, Libera Università Maria Ss. Assunta, Roma, Italia \\
           Department of Mathematics, University of Bergen, Bergen, Norway\\
}

\date{}

\maketitle

\begin{abstract}
Hidden Markov models (HMMs) have been extensively used in the univariate and multivariate literature.
However, there has been an increased interest in the analysis of matrix-variate data over the recent years.
In this manuscript we introduce HMMs for matrix-variate longitudinal data, by assuming a matrix normal distribution in each hidden state.
Such data are arranged in a four-way array.
To address for possible overparameterization issues, we consider the spectral decomposition of the covariance matrices, leading to a total of 98 HMMs.
An expectation-conditional maximization algorithm is discussed for parameter estimation.
The proposed models are firstly investigated on simulated data, in terms of parameter recovery, computational times and model selection.
Then, they are fitted to a four-way real data set concerning the unemployment rates of the Italian provinces, evaluated by gender and age classes, over the last 16 years.

\keywords{Hidden Markov models \and Matrix-variate \and Clustering \and Parsimonious models}
\end{abstract}

\section{Introduction}
\label{sec:intro}

Multivariate longitudinal data have been widely analyzed in the literature (\citealp{Verbeke2014} and \citealp{Verdam2019}).
Data are usually presented in the standard three-way format, where units, times and variables are arranged in software-ready manners.
Because of their three-way structure, multivariate longitudinal data have been recently arranged in a matrix-variate fashion (\citealp{huang2019} and \citealp{viroli2011model}): for each unit $i=1,\dots,I$, we observe a $P \times T$ matrix, where $P$ and $T$ denote the number of variables and times, respectively. 
Then, such data have been used for model-based clustering via matrix-variate mixture models (see e.g.~\citealp{Wang2020,melnykov2019studying,tomarchio2020mixtures,tomarchio2020two}).
This allows for both clustering units in homogeneous groups, defined according to similarities between matrix-variate data, and separately modeling the association between variables and times. 	
Unfortunately, this procedure has two side effects:
\begin{enumerate}[(a)] 
\item using the time on either the rows or the columns of the matrices reduces the types of longitudinal data structures that can be arranged in a matrix-variate framework. 
For instance, spatio-temporal data are used to either to analyze $P$ variables observed at $T$ times for $R$ different locations \citep{viroli2011model} or to evaluate one measurement on $R$ locations at $T$ times on a set of $I$ units \citep{viroli2011finite}.
However, it is not possible to jointly consider $P$ variables at $R$ locations for $T$ times on $I$ units.
A possible solution could be to combine locations-times in one $RT$-dimension, as done by \citet{viroli2011finite}, but this implies a loss in terms of interpretability as well as an increase in the number of parameters of the models, given the higher dimensionality of the matrices.

Another example consists of two-factor data, which have been commonly considered in longitudinal settings (see e.g.~\citealp{brunner2001}, \citealp{fitzmaurice2008primer}, \citealp{noguchi2012}). 
Such data, have been recently used in matrix-variate mixture models by \citep{sarkar2020parsimonious} in a not-longitudinal way, given that the factors fill the two dimensions of the $P \times R$ matrices for the $I$ units, and an additional dimension for the time is required.
In other terms, in both examples it would be necessary to move from three-way to four-way data structures.
 
\item Currently available clustering approaches for matrix-variate data assume time-constant clustering, i.e.~it is not possible for the sample units to move across clusters over time and the evolution over time of the clustering structure is completely overlooked. \cite{zhu2021finite} recognize the importance of allowing for time-dependence in a matrix-variate clustering setting and propose a procedure capable of capturing the heterogeneity pattern and estimating change points from all data groups simultaneously. 
Indeed, time-varying heterogeneity is a specific important feature of longitudinal data analysis, and as such appropriate modeling strategies should be considered. 
Hidden Markov models (HMMs) have been extensively used to address this longitudinal data peculiarity (\citealp{zucchini2017hidden}, \citealp{bartolucci2012latent}, \citealp{maruotti2011mixed} and \citealp{alt:2007}). 
Being (dependent) mixtures, HMMs simultaneously allow for clustering units and for modeling the evolution of the clustering over time.
\end{enumerate}

To jointly consider the aspects in (a) and (b), in this manuscript we introduce and discuss HMMs for matrix-variate longitudinal data, with a specific application on the two-factor longitudinal case.
Such kind of data can be arranged in a four-way array of dimension $P \times R \times I \times T$.
Under this four-way longitudinal data setting, HMMs simultaneously investigates several further (null) hypotheses, beyond the recovering of the time-varying clustering:
\begin{itemize}
	\item all means at different time points are the same; 
	\item there is an association between the factor levels.
\end{itemize} 
The association is often the primary interest when researchers aim to study whether levels differentially affect outcomes in different hidden states.

A side effect of working with four-way data is the large number of parameters involved.
This often occurs because of the (row- and column-specific) covariance matrices, since $P(P+1)/2$ and $R(R+1)/2$ unique parameters must be estimated.
One of the most classical ways of addressing this overparameterization issue involves the spectral decomposition of the covariance matrices \citep{celeux1995gaussian}.
This decomposition offers remarkable flexibility and a geometric interpretation in terms of volume, shape, and orientation of the hidden states (for other approaches available in the HMMs literature see \citealp{Maruotti2017} and \citealp{farcomeni2020dimension}).
By using the spectral decomposition of the covariance matrices, we obtain a family of 98 parsimonious HMMs that will be described in Section~\ref{subsec:pars}, after the presentation of the general model (Section~\ref{subsec:gen}).
In this framework, model parameters can be estimated by a full maximum likelihood method based on the Expectation Conditional Maximization (ECM) algorithm \citep{meng1993maximum}, and recursions widely used in the HMM literature \citep{baum1970maximization}.
An iterative Minorization–Maximization (MM) algorithm \citep{browne2014} is also adopted to estimate a subset of the parsimonious HMMs.

We illustrate the proposal by a large-scale simulation study in order to investigate the empirical behavior of the proposed approach with respect to several aspects, such as the number of observed units and times, the number of hidden states and the association structure between factor-levels. 
We focus on goodness of clustering and parameters recovery, with a focus on computational times and model selection procedures. 
Furthermore, we test the proposal by analyzing a sample taken from the Italian National Institute of Statistics on the unemployment rate in 98 Italian provinces recorded for 16 years, also covering the 2008 crisis. 
We examine the unemployment rate arranged as a two-factor design, i.e.~taking into account gender and age classes, by allowing some dynamics in the evolution of unemployment. 
We obtain a flexible model by including different associations across levels, changing according to the inferred dynamics, and by accounting for unobserved characteristics influencing changes in the province's unemployment patterns.

Finally, Section~\ref{sec:conc} summarizes the key aspects of our proposal along with future possible extensions.

\section{Methodology}
\label{sec:meth}

\subsection{The model}
\label{subsec:gen}

Let $\left\{\calX_{it}; i=1,\ldots,I, t=1,\ldots,T\right\}$ be a sequence of matrix-variate longitudinal observations recorded on $I$ units over $T$ times, with $\calX_{it}\in \Rset^{R\times P}$, and let $\left\{S_{it}; i=1,\ldots,I, t=1,\ldots,T\right\}$ be a first-order Markov chain defined on the state space $\left\{1,\ldots,k,\ldots,K\right\}$.
As mentioned in Section~\ref{sec:intro}, a HMM is a particular type of dependent mixture model consisting of two parts: an underlying unobserved process $\left\{S_{it}\right\}$ that satisfies the Markov property, i.e.
\begin{equation*}
\text{Pr}\left(S_{it}=s_{it} | S_{i1}=s_{i1}, \ldots,S_{it-1}=s_{it-1}\right)=\text{Pr}\left(S_{it}=s_{it} | S_{it-1}=s_{it-1}\right),
\label{eq:1}
\end{equation*}
and a state-dependent observation process $\left\{\calX_{it}\right\}$ for which the conditional independence property holds, i.e.
\begin{align*}
& f\left(\calX_{it}=\vecX_{it} | \calX_{i1}=\vecX_{i1}, \ldots,\calX_{it-1}=\vecX_{it-1},S_{i1}=s_{i1}\ldots,S_{it}=s_{it}\right) \nonumber \\
& = f\left(\calX_{it}=\vecX_{it} | S_{it}=s_{it} \right),
\label{eq:2}
\end{align*}
where $f(\cdot)$ is a generic probability density function (pdf).
Therefore, the unknown parameters in an HMM involve both the parameters of the Markov chain and those of the state-dependent pdfs.
In detail, the parameters of the Markov chain are the initial probabilities $\pi_{ik}=\text{Pr}\left(S_{i1}=k\right)$, $k=1,\ldots,K$, being $K$ the number of states, and the transition probabilities
\begin{equation*}
\pi_{ik|j}=\text{Pr}\left(S_{it}=k|S_{it-1}=j\right), \hspace{0.5em} t=2,\ldots,T  \hspace{0.5em} \text{and} \hspace{0.5em} j,k=1,\ldots,K,
\label{eq:3}
\end{equation*}
where $k$ refers to the current state and $j$ refers to the one previously visited.
To simplify the discussion, we will consider homogeneous HMMs, that is $\pi_{ik|j}=\pi_{k|j}$ and $\pi_{ik}=\pi_{k}, i=1,\ldots,I$.
We collect the initial probabilities in the $K$-dimensional vector $\bpi$, whereas the time-homogenous transition probabilities are inserted in the $K \times K$ transition matrix $\bPi$.

Regarding the conditional density for the observed process, it will be given by a matrix-normal distribution, i.e.
\begin{equation}
\phi\left(\vecX_{it}|S_{it}=k;\vectheta_{k}\right) = \frac{\exp\left\{-\frac{1}{2}\tr\left[\matsig_{k}^{-1}(\vecX-\matm_{k})\matPsi_{k}^{-1}(\vecX-\matm_{k})'\right]\right\}}{(2\pi)^{\frac{PR}{2}}|\matsig_{k}|^{\frac{R}{2}}|\matPsi_{k}|^{\frac{P}{2}}},
\label{eq:matnorm}
\end{equation}
where $\matm_{k}$ is the $P \times R$ matrix of means, $\matsig_{k}$ is the $P\times P$ covariance matrix containing the variances and covariances between the P rows, $\matPsi_{k}$ is the $R\times R$ covariance matrix containing the variance and covariances of the $R$ columns and $\vectheta_{k}=\left\{\matm_{k},\matsig_{k},\matPsi_{k}\right\}$.
For an exhaustive description of the matrix-normal distribution and its properties see \citet{gupta2018matrix}.

\subsection{Parsimonious models}
\label{subsec:pars}

As introduced in Section~\ref{sec:intro}, a way to reduce the number of parameters of the model is to introduce parsimony in the covariance matrices via the well-known eigen decomposition.
Specifically, a $Q \times Q$ covariance matrix can be decomposed as
\begin{equation}
\bPhi_{k} = \lambda_{k}\bGamma_{k}\bDelta_{k}\bGamma_{k}'
\label{eq:decomp}
\end{equation}
where $\lambda_{k}=|\bPhi_{k}|^{1/Q}$, $\bGamma_{k}$ is a $Q \times Q$ orthogonal matrix of the eigenvectors of $\bPhi_{k}$ and $\bDelta_{k}$ is a diagonal matrix with the eigenvalues of $\bPhi_{k}$ located on the main diagonal.
From a geometric point of view, $\lambda_{k}$ determines the volume, $\bGamma_{k}$ indicates the orientation, and $\bDelta_{k}$ denotes the shape of the $k$th state.
By imposing constraints on the three components of~\eqref{eq:decomp}, the fourteen parsimonious models of Table~\ref{tab:parsi} are obtained.
\begin{table}[!htt]
	\caption{Nomenclature, covariance matrix structure, and number of free parameters in $\bPhi_{1},\ldots,\bPhi_{K}$ for the
parsimonious models obtained via the eigen decomposition of the state covariance matrices.}
\resizebox{\textwidth}{!}{%
	\centering
	\begin{tabular}{lllllll}
				\hline\noalign{\smallskip}
		Family & Model & Type & Volume & Shape & Orientation & \# of free parameters \\
		 &     &      &        &       &             &  in $\bPhi_{1},\ldots,\bPhi_{K}$ \\
	      \noalign{\smallskip}\hline\noalign{\smallskip}
		Spherical & EII & $\lambda\bI$                                        &  Equal    & Spherical & - & 1 \\
		Spherical & VII & $\lambda_{k}\bI$                                    &  Variable & Spherical & - & $K$ \\[1.2mm]
		Diagonal & EEI & $\lambda\bDelta$                                     &  Equal    & Equal & Axis-Aligned & $Q$\\
		Diagonal & VEI & $\lambda_{k}\bDelta$                                 &  Variable & Equal & Axis-Aligned & $K + Q - 1$\\
		Diagonal & EVI & $\lambda\bDelta_{k}$                                 &  Equal    & Variable & Axis-Aligned & $K(Q-1)+1$\\
		Diagonal & VVI & $\lambda_{k}\bDelta_{k}$                             &  Variable & Variable & Axis-Aligned & $KQ$\\[1.2mm]
		General  & EEE & $\lambda\bGamma\bDelta\bGamma^{\top}$                &  Equal    & Equal & Equal & $Q(Q+1)/2$\\
		General & VEE & $\lambda_{k}\bGamma\bDelta\bGamma^{\top}$             &  Variable & Equal & Equal & $Q(Q+1)/2 + K - 1$\\
		General & EVE & $\lambda\bGamma\bDelta_{k}\bGamma^{\top}$             &  Equal    & Variable & Equal & $Q(Q-1)/2 + K(Q-1)+1 $ \\
		General & VVE & $\lambda_{k}\bGamma\bDelta_{k}\bGamma^{\top}$         &  Variable & Variable & Equal & $Q(Q-1)/2 + KQ$\\
		General & EEV & $\lambda\bGamma_{k}\bDelta\bGamma_{k}^{\top}$         &  Equal    & Equal & Variable & $KQ(Q-1)/2 + Q$\\
		General & VEV & $\lambda_{k}\bGamma_{k}\bDelta\bGamma_{k}^{\top}$     &  Variable & Equal & Variable & $KQ(Q-1)/2 + K + Q - 1$\\
		General & EVV & $\lambda\bGamma_{k}\bDelta_{k}\bGamma_{k}^{\top}$     &  Equal    & Variable & Variable & $KQ(Q+1)/2 -K +1$\\
		General & VVV & $\lambda_{k}\bGamma_{k}\bDelta_{k}\bGamma_{k}^{\top}$ &  Variable & Variable & Variable & $KQ(Q+1)/2$\\
		\noalign{\smallskip}\hline
	\end{tabular}%
}	
\label{tab:parsi}
\end{table}

Considering that we have two covariance matrices in~\eqref{eq:matnorm}, this would yield to $14 \times 14 = 196$ parsimonious HMMs.
However, there is a non-identifiability issue since $\matPsi \otimes \matsig = \matPsi^{*} \otimes \matsig^{*}$ if $\matsig^{*}= a\matsig$ and $\matPsi^{*}= a^{-1}\matPsi$.
As a result, $\matsig$ and $\matPsi$ are identifiable up to a multiplicative constant $a$ \citep{sarkar2020parsimonious}.
To avoid such issue, the column covariance matrix $\matPsi$ is restricted to have $|\matPsi|= 1$, implying that in~\eqref{eq:decomp} the parameter $\lambda_{k}$ is unnecessary. 
This reduces the number of models related to $\matPsi$ from 14 to 7, i.e., $\bI, \bDelta, \bDelta_{k}, \bGamma\bDelta\bGamma^{\top},\bGamma\bDelta_{k}\bGamma^{\top}, \bGamma_{k}\bDelta\bGamma_{k}^{\top}, \bGamma_{k}\bDelta_{k}\bGamma_{k}^{\top}$.
Therefore, we obtain $14 \times 7 = 98$ parsimonious HMMs. 

\subsection{Maximum likelihood estimation}
\label{subsec:llk}

To fit our HMMs, we use the expectation-conditional maximization (ECM) algorithm \citep{meng1993maximum}.
The ECM algorithm is a variant of the classical expectation-maximization (EM) algorithm \citep{dempster77}, from which it differs since the M-step is replaced by a sequence of simpler and computationally convenient CM-steps.

Let $S=\left\{\vecX_{it}; i=1,\ldots,I, t=1,\ldots,T\right\}$ be a sample of matrix-variate longitudinal observations.
Then, the incomplete-data likelihood function is 
\begin{equation*}
L\left(\vecTheta|S\right) = \prod_{i=1}^{I} \bpi' \bphi\left(\vecX_{i1}\right) \bPi \bphi\left(\vecX_{i2}\right) \bPi \ldots \bphi\left(\vecX_{iT-1}\right) \bPi \bphi\left(\vecX_{iT}\right) \bone_K,
\label{eq:lk}
\end{equation*}
where $\bphi\left(\vecX_{it}\right)$ is a $K \times K$ diagonal matrix with conditional densities $\calX_{it}=\vecX_{it} | S_{it}=k$ on the main diagonal, $\bone_K$ is a vector $K$ ones and $\vecTheta$ contains all the model parameters.
In this setting, $\Sa$ is viewed as incomplete because, for each observation, we do not know its state membership and its evolution over time.
For this reason, let us define the unobserved state membership $\bz_{it}=\left(z_{it1},\ldots,z_{itk},\ldots,z_{itK}\right)'$ and the unobserved states transition
$$\bz\bz_{it}= \begin{bmatrix}
           zz_{it11} & \ldots & zz_{it1k} & \ldots & zz_{it1K}\\
           \vdots    &        & \vdots    &        & \vdots   \\   
           zz_{itj1} & \ldots & zz_{itjk} & \ldots & zz_{itjK}\\
					 \vdots    &        & \vdots    &        & \vdots   \\   
           zz_{itK1} & \ldots & zz_{itKk} & \ldots & zz_{itKK}\end{bmatrix}$$
where
$$ z_{itk}= \begin{cases}
1 & \quad \text{if}\hspace{0.5em} S_{t}=k \\
0 & \quad \text{otherwise}
\end{cases} \text{and}
\hspace{0.5em}
zz_{itjk}= \begin{cases}
1 & \quad \text{if} \hspace{0.5em} S_{it-1}=j \hspace{0.5em}\text{and}\hspace{0.5em} S_{it}=k \\
0 & \quad \text{otherwise}
\end{cases}.
$$					
Therefore, the complete data are $\Sa_c=\left\{\vecX_{it},\bz_{it},\bz\bz_{it}; i=1,\ldots,I, t=1,\ldots,T\right\}$
and the corresponding complete-data log-likelihood is
\begin{equation}
l_{c}\left(\vecTheta|\Sa_{c}\right) = l_{c_1}\left(\bpi|\Sa_{c}\right)+l_{c_2}\left(\bPi|\Sa_{c}\right)+l_{c_3}\left(\vectheta|\Sa_{c}\right),
\label{eq:llk}
\end{equation}
with $\vectheta=\left\{\vectheta_k; k=1,\ldots,K\right\}$ and 
\begin{align*}
l_{c_1}\left(\bpi|\Sa_{c}\right) & = \sum\limits_{i=1}^{I}\sum\limits_{k=1}^{K} z_{i1k} \log\left(\pi_k\right) \\
l_{c_2}\left(\bPi|\Sa_{c}\right) & = \sum\limits_{i=1}^{I}\sum\limits_{t=2}^{T}\sum\limits_{k=1}^{K}\sum\limits_{j=1}^{K} zz_{itjk} \log\left(\pi_{k|j}\right) \\
l_{c_3}\left(\vectheta|\Sa_{c}\right) & = \sum\limits_{i=1}^{I}\sum\limits_{t=1}^{T}\sum\limits_{k=1}^{K} z_{itk}\Bigg[\Bigg.-\frac{PR}{2}\log\left(2\pi\right)-\frac{R}{2}\log|\matsig_{k}|-\frac{P}{2}\log|\matPsi_{k}| \\
& -\frac{1}{2}\tr\left[\matsig_{k}^{-1}(\vecX-\matm_{k})\matPsi_{k}^{-1}(\vecX-\matm_{k})'\right]\Bigg. \Bigg].
\end{align*}
In the following, by adopting the notation used in \citet{melnykov2019studying}, the parameters marked with one dot represent the updates at the previous iteration and those marked
with two dots are the updates at the current iteration.

\paragraph{E-Step} 
\label{par:e-st}

The E-step requires calculation of the conditional expectation of~\eqref{eq:llk}, given $\Sa_{c}$ and the current estimates of $\dot{\vecTheta}$.
Therefore, we need to replace $z_{itk}$ and $z_{itjk}$ with their conditional expectations, namely, $\ddot{z}_{itk}$ and $\ddot{zz}_{itjk}$.
This can be efficiently done by exploiting a forward recursion approach \citep{baum1970maximization, zucchini2017hidden}.

Let us start by defining the forward probability
\begin{equation*}
\gamma_{itk}=\text{Pr}\left(\calX_{i1}=\vecX_{i1},\ldots,\calX_{it}=\vecX_{it},S_{it}=k\right),
\label{eq:fw}
\end{equation*}
that is the probability of seeing the partial sequence finishing up in state $k$ at time $t$, and the corresponding backward probability
\begin{equation*}
\beta_{itk}=\text{Pr}\left(\calX_{it+1}=\vecX_{it+1},\ldots,\calX_{iT}=\vecX_{iT}|S_{it}=k\right).
\label{eq:bw}
\end{equation*}
It is known that the computation of the forward and backward probabilities is susceptible to numerical overflow errors \citep{zucchini2017hidden}.
To prevent, or at least to decrease, the risk of such errors a scaling procedure is used.
Specifically, it will be convenient to work on the log-scale \citep{farcomeni2012}.
In order to do so, we use a simple computational device based on the following equality:
$$\log\left(a+b\right)=\log(a)+\log\left(1+\exp\left(\log\left(b\right)-\log\left(a\right)\right)\right).$$
Therefore, if one has only the log of two quantities $\log\left(a\right)$ and $\log\left(b\right)$, only their difference must be exponentiated to obtain $\log\left(a+b\right)$, reducing the risks of underflow.
By iterating this reasoning, one can sum a vector of quantities on the log-scale.
This operation is called $\bigoplus$.
Thus, when $t=1$, the forward recursion on the log-scale is given by
\begin{equation*}
\log\left(\gamma_{i1k}\right) = \log\left[\phi\left(\vecX_{i1}|S_{i1}=k\right)\right]+\log\left(\pi_k\right)
\label{eq:fw1}
\end{equation*}
whereas, for $t=2,\ldots,T$, it is
\begin{equation*}
\log\left(\gamma_{itk}\right) = \log\left[\phi\left(\vecX_{it}|S_{it}=k\right)\right]+\log\left(\pi_k|j\right)+\bigoplus_{j=1}^K \log\left(\gamma_{it-1,j}\right).
\label{eq:fw2}
\end{equation*}
In a similar way, for the backward recursion on the log-scale is
$$\log\left(\beta_{iTk}=0\right),$$
and, for $t=T-1,\ldots,1$, we obtain
\begin{equation*}
\log\left(\beta_{itj}\right) = \bigoplus_{k=1}^K \log\left[\phi\left(\vecX_{it+1}|S_{it+1}=k\right)\right]+\log\left[\beta_{it+1,k}+\log\left(\pi_{k|j}\right)\right].
\label{eq:bw1}
\end{equation*}
Then, the updates required in the E-step can be computed as 
\begin{equation*}
\ddot{z}_{itk} = \frac{\gamma_{itk}\beta_{itk}}{\sum\limits_{h=1}^{K}\gamma_{ith}\beta_{ith}}, \quad \ddot{zz}_{itjk} = \frac{\gamma_{i\left(t-1\right)j}\pi_{k|j}\phi\left(\vecX_{it}|S_{it}=k\right)\beta_{itk}}{\sum\limits_{h=1}^{K}\gamma_{iTh}}.
\label{eq:e1}
\end{equation*}

\paragraph{CM-Step 1} 
\label{par:cm1}

Consider $\vecTheta=\left\{\vecTheta_1,\vecTheta_2\right\}$, where $\vecTheta_1=\left\{\pi_k,\bPi,\matm_{k},\matsig_{k}\right\}$ and $\vecTheta_2=\left\{\matPsi_{k}\right\}$.
At the first CM-step, we maximize the expectation of the complete-data log-likelihood with respect to $\vecTheta_1$, fixing $\vecTheta_2$ at $\dot{\vecTheta_2}$.
In particular, we obtain
\begin{equation*}
\ddot{\pi}_k    = \frac{\sum_{i=1}^I \ddot{z}_{i1k}}{I}, \quad \ddot{\pi}_{k|j} = \frac{\sum_{i=1}^I \sum_{t=2}^T \ddot{zz}_{itjk}}{\sum_{i=1}^I \sum_{t=2}^T \sum_{k=1}^K \ddot{zz}_{itjk}},
\label{eq:cm2}
\end{equation*}
\begin{equation*}
\ddot{\matm}_k   = \frac{\sum_{i=1}^I \sum_{t=1}^T \ddot{z}_{itk}\vecX_{it}}{\sum_{i=1}^I \sum_{t=1}^T \ddot{z}_{itk}}.
\label{eq:cm3} 
\end{equation*}
The update for $\matsig_k$ depends on the parsimonious structure considered.
For notational simplicity, let $\ddot{\vecY}=\sum_{k=1}^K \ddot{\vecY}_k$ be the update of the within state row scatter matrix, where $\ddot{\vecY}_k = \sum_{i=1}^I \sum_{t=1}^T \ddot{z}_{itk}\left(\vecX_{it}-\ddot{\matm}_k\right)\dot\matPsi_k^{-1}\left(\vecX_{it}-\ddot{\matm}_k\right)'$ is the update of the row scatter matrix related to the $k$th state.
The updates for the 14 parsimonious structures of $\matsig_k$ are:
\begin{itemize}

\item Model EII [$\matsig_k=\lambda\bI$] In this setting, the row covariance matrices of all states are spherical and have equal volume. We need to estimate only $\lambda$ as
\begin{equation*}
\ddot\lambda = \frac{\tr\left\{\ddot \vecY\right\}}{PRTI}.
\label{eq:eii}
\end{equation*}

\item Model VII [$\matsig_k=\lambda_{k}\bI$] In this case, the row covariance matrices are spherical but their volume is different. Thus, the update for $\lambda_{k}$ is
\begin{equation*}
\ddot\lambda_k = \frac{\tr\left\{\ddot \vecY_k\right\}}{PR \sum_{i=1}^I \sum_{t=1}^T \ddot{z}_{itk}}.
\label{eq:vii}
\end{equation*}

\item Model EEI [$\matsig_k=\lambda\bDelta$] Here, the row covariance matrices of all states have equal volume, shape and are axis-aligned. The updates for $\bDelta$ and $\lambda$ are
\begin{equation*}
\ddot\bDelta =\frac{\text{diag}\left(\ddot \vecY\right)}{ \left|\text{diag}\left(\ddot \vecY\right) \right|^\frac{1}{P}}, \quad \ddot\lambda = \frac{\left| \text{diag}\left(\ddot \vecY\right)\right|^\frac{1}{P}}{RTI}.
\label{eq:eei}
\end{equation*}

\item Model VEI [$\matsig_k=\lambda_{k}\bDelta$] In this setting, the row covariance matrices of all states have equal shape and are axis-aligned, but they are allowed to have different volumes.
We need to update $\bDelta$ and $\lambda$ as
\begin{equation*}
\ddot\bDelta = \frac{\text{diag}\left(\sum\limits_{k=1}^K \dot \lambda_k^{-1}\ddot \vecY_k\right)}{\left|\text{diag}\left(\sum\limits_{k=1}^K \dot\lambda_k^{-1}\ddot \vecY_k\right)\right|^\frac{1}{P}}, \quad  \ddot\lambda_k = \frac{\tr\left\{\ddot \vecY_k \ddot\bDelta^{-1}\right\}}{PR\sum_{i=1}^I \sum_{t=1}^T \ddot{z}_{itk}}.
\label{eq:vei}
\end{equation*}

\item Model EVI [$\matsig_k=\lambda\bDelta_{k}$] In this case, the row covariance matrices have equal volume and are axis-aligned, but they have different shapes. The updates for $\bDelta_{k}$ and $\lambda$ are
\begin{equation*}
\ddot\bDelta_k = \frac{\text{diag}\left(\ddot \vecY_k\right)}{\left|\text{diag}\left(\ddot \vecY_k\right)\right|^\frac{1}{P}}, \quad  \ddot\lambda = \frac{\sum\limits_{k=1}^K\left|\text{diag}\left(\ddot \vecY_k\right)\right|^\frac{1}{P}}{RTI}.
\label{eq:evi}
\end{equation*}

\item Model VVI [$\matsig_k=\lambda_{k}\bDelta_{k}$] The most generic case in the diagonal family has row covariance matrices with different volume and shape, in addition to being axis-aligned.
Then, the updates for $\bDelta_{k}$ and $\lambda_{k}$ are
\begin{equation*}
\ddot\bDelta_k = \frac{\text{diag}\left(\ddot \vecY_k\right)}{\left|\text{diag}\left(\ddot \vecY_k\right)\right|^\frac{1}{P}}, \quad  \ddot\lambda_k = \frac{\left|\text{diag}\left(\ddot \vecY_k\right)\right|^\frac{1}{P}}{D\sum_{i=1}^I \sum_{t=1}^T \ddot{z}_{itk}}.
\label{eq:vvi}
\end{equation*}

\item Model EEE [$\matsig_k=\lambda\bGamma\bDelta\bGamma^{\top}$] The most constrained member of the general family has row covariance matrices with same volume, shape and orientation.
Thus, the update for $\matsig$ is given by
\begin{equation*}
\ddot{\matsig}= \frac{\ddot \vecY}{RTI}. 
\label{eq:eee}
\end{equation*}

\item Model VEE [$\matsig_k=\lambda_{k}\bGamma\bDelta\bGamma^{\top}$] This model assumes row covariances matrices with same shape and orientation but different volumes. Let $\vecC = \ddot\bGamma\ddot\bDelta\ddot\bGamma^{\top}$. The updates for $\vecC$ and $\lambda_{k}$ are
\begin{equation*}
\ddot \vecC = \frac{\sum\limits_{k=1}^K \dot\lambda_k^{-1}\ddot \vecY_k}{\left|\sum\limits_{k=1}^K \dot\lambda_k^{-1}\ddot \vecY_k\right|^{\frac{1}{P}}}, \quad  \ddot\lambda_k = \frac{\tr\left\{\ddot \vecC^{-1} \ddot\vecY_k \right\}}{PR\sum_{i=1}^I \sum_{t=1}^T \ddot{z}_{itk}}. 
\label{eq:vee}
\end{equation*}

\item Model EVE [$\matsig_k=\lambda\bGamma\bDelta_{k}\bGamma^{\top}$] Here, the row covariance matrices have equal volume and orientation but different shapes.
Given that there is no analytical solution for $\bGamma$, while keeping fixed the other parameters, an iterative Minorization–Maximization (MM) algorithm \citep{browne2014} is employed.
In detail, a surrogate function can be constructed as 
\begin{equation*}
f\left(\bGamma\right) = \sum\limits_{k=1}^K \tr\left\{\vecY_k\bGamma\bDelta_{k}^{-1}\bGamma^{\top}\right\} \leq S + \tr\left\{\bF\bGamma\right\},
\label{eq:mm1}
\end{equation*}
where $S$ is a constant and $\bF = \sum_{k=1}^K\left(\bDelta_{k}^{-1} \dot\bGamma^{\top} \vecY_k - e_k \bDelta_{k}^{-1} \dot\bGamma^{\top}\right)$, with $e_k$ being the largest eigenvalue of $\vecY_k$.
The update of $\bGamma$ is given by $\ddot \bGamma = \dot \bG \dot \bH ^{\top}$, where $\dot \bG$ and $\dot \bH$ are obtained from the singular value decomposition of $\bF$.
This process is repeated until a specified convergence criterion is met and the estimate $\ddot \bGamma$ is obtained from the last iteration.
Then, we obtain the update for $\bDelta_{k}$ and $\lambda$ as
\begin{equation*}
\ddot\bDelta_k = \frac{\text{diag}\left(\ddot\bGamma^{\top} \ddot \vecY_k \ddot\bGamma\right)}{\left|\text{diag}\left(\ddot\bGamma^{\top} \ddot \vecY_k \ddot\bGamma\right)\right|^\frac{1}{P}}, \quad  \ddot\lambda = \frac{\sum\limits_{k=1}^K \tr\left(\ddot\bGamma \ddot\bDelta_k^{-1} \ddot\bGamma^{\top}\ddot \vecY_k\right)}{PRTI}. 
\label{eq:eve}
\end{equation*}

\item Model VVE [$\matsig_k=\lambda_{k}\bGamma\bDelta_{k}\bGamma^{\top}$] In this case, the row covariance matrices have the same orientation, but varying volumes and shapes.
Again, there is no analytical solution for $\bGamma$, and its update is obtained by employing the MM algorithm as described for the EVE model.
Then, the updates for $\bDelta_{k}$ and $\lambda_{k}$ are
\begin{equation*}
\ddot\bDelta_k = \frac{\text{diag}\left(\ddot\bGamma^{\top} \ddot \vecY_k \ddot\bGamma\right)}{\left|\text{diag}\left(\ddot\bGamma^{\top} \ddot \vecY_k \ddot\bGamma\right)\right|^\frac{1}{P}}, \quad  \ddot\lambda_k = \frac{\left|\text{diag}\left(\ddot\bGamma^{\top} \ddot \vecY_k \ddot\bGamma\right)\right|^{\frac{1}{P}}}{R\sum_{i=1}^I \sum_{t=1}^T \ddot{z}_{itk}}. 
\label{eq:vve}
\end{equation*}

\item Model EEV [$\matsig_k=\lambda\bGamma_{k}\bDelta\bGamma_{k}^{\top}$] Here, row covariance matrices have the same volume and shape, but different orientations.
An algorithm similar to the one proposed by \citet{celeux1995gaussian} can here employed.
In detail, the eigen-decomposition $\vecY_k=\bL_k \vecOmega_k \bL_k^{\top}$ is firstly considered, with eigenvalues in the diagonal matrix $\vecOmega_k$ following descending order and orthogonal matrix $\bL_k$ composed of the corresponding eigenvectors.
Then, we obtain the update for $\bGamma_k$, $\bDelta$ and $\lambda$ as
\begin{equation*}
\ddot\bGamma_k=\ddot \bL_k , \quad \ddot\bDelta = \frac{\sum\limits_{k=1}^K \ddot \vecOmega_k}{\left|\sum\limits_{k=1}^K \ddot \vecOmega_k\right|^\frac{1}{P}}, \quad  \ddot\lambda = \frac{\left|\sum\limits_{k=1}^K \ddot \vecOmega_k\right|^\frac{1}{P}}{RTI}. 
\label{eq:eev}
\end{equation*}

\item Model VEV [$\matsig_k=\lambda_k\bGamma_{k}\bDelta\bGamma_{k}^{\top}$] In this setting, row covariance matrices have the same shape, but different volumes and orientations.
By using the same algorithm applied for the EEV model, the update for $\bGamma_k$, $\bDelta_{k}$ and $\lambda_{k}$ are
\begin{equation*}
\ddot\bGamma_k=\ddot \bL_k , \quad \ddot\bDelta = \frac{\sum\limits_{k=1}^K \lambda_k^{-1} \ddot \vecOmega_k}{\left|\sum\limits_{k=1}^K \lambda_k^{-1} \ddot \vecOmega_k\right|^\frac{1}{P}}, \quad  \ddot\lambda_k = \frac{\tr\left\{\ddot \vecOmega_k \ddot\bDelta^{-1} \right\}}{PR\sum_{i=1}^I \sum_{t=1}^T \ddot{z}_{itk}}. 
\label{eq:vev}
\end{equation*}

\item Model EVV [$\matsig_k=\lambda\bGamma_{k}\bDelta_{k}\bGamma_{k}^{\top}$] In this model, row covariance matrices have varying shapes and orientations but equal volume.
The updates of this model can be obtained in a similar fashion of the EVI model.
Thus, by considering $\vecC_k = \bGamma_k\bDelta_k\bGamma_k^{\top}$, we estimate $\bGamma_k$, $\bDelta_{k}$ and $\lambda$ as
\begin{equation*}
\ddot \vecC_k = \frac{\ddot \vecY_k}{\left|\ddot \vecY_k\right|^{\frac{1}{P}}}, \quad  \ddot\lambda = \frac{\sum\limits_{k=1}^K \left|\ddot \vecY_k\right|^{\frac{1}{P}}}{RTI}. 
\label{eq:evv}
\end{equation*}

\item Model VVV [$\matsig_k=\lambda_{k}\bGamma_{k}\bDelta_{k}\bGamma_{k}^{\top}$] In the full unconstrained case, we obtain 
\begin{equation*}
\ddot \matsig_k = \frac{\ddot \vecY_k}{R\sum_{i=1}^I \sum_{t=1}^T \ddot{z}_{itk}}.
\label{eq:vvv}
\end{equation*}

\end{itemize}
\
\paragraph{CM-Step 2} 
\label{par:cm2}

At the second CM-step, we maximize the expectation of the complete-data log-likelihood with respect to $\vecTheta_{2}$, keeping $\vecTheta_{1}$ fixed at $\ddot\vecTheta_{1}$.
The update for $\matPsi_k$ depends on which of the 7 parsimonious structure is considered.
For notational simplicity, let $\ddot{\vecW}=\sum_{k=1}^K \ddot{\vecW}_k$ be the update of the within state column scatter matrix, where $\ddot{\vecW}_k = \sum_{i=1}^I \sum_{t=1}^T \ddot{z}_{itk}\left(\vecX_{it}-\ddot{\matm}_k\right)'\ddot \matsig_k^{-1}\left(\vecX_{it}-\ddot{\matm}_k\right)$ is the update of the column scatter matrix related to the $k$th state.
In detail, we have:

\begin{itemize}
\item Model II [$\matPsi_k=\bI$] This is the simpler model, since the column covariance matrices are spherical and assumed to be $R \times R$ identity matrices. Therefore, there are no parameters to be estimated.

\item Model EI [$\matPsi_k=\bDelta$] In this setting, the column covariance matrices have the same shape and are axis-aligned.
The update for $\bDelta$ is  
\begin{equation*}
\ddot \bDelta = \frac{\text{diag}\left(\ddot \vecW\right)}{\left|\text{diag}\left(\ddot \vecW\right) \right|^\frac{1}{R}}.
\label{eq:ei}
\end{equation*}

\item Model VI [$\matPsi_k=\bDelta_k$] Here, the column covariance matrices have different shapes and are axis-aligned.
Thus, we update $\bDelta_k$ as 
\begin{equation*}
\ddot \bDelta_k = \frac{\text{diag}\left(\ddot \vecW_k\right)}{\left|\text{diag}\left(\ddot \vecW_k\right) \right|^\frac{1}{R}}.
\label{eq:vi}
\end{equation*}

\item Model EE [$\matPsi_k=\bGamma\bDelta\bGamma^{\top}$] In this case, the column covariance matrices have equal shapes and orientations.
Therefore, we can directly obtain  
\begin{equation*}
\ddot \matPsi = \frac{\ddot \vecW}{\left|\ddot \vecW \right|^\frac{1}{R}}.
\label{eq:ee}
\end{equation*}

\item Model VE [$\matPsi_k=\bGamma\bDelta_k\bGamma^{\top}$] In this setting, the column covariance matrices have equal orientation but different shapes.
Similarly to the EVE and VVE models in CM-Step 1, there is no analytical solution for $\bGamma$, while keeping fixed the other parameters.
Therefore, the MM algorithm is implemented by following the same procedure explained for the EVE model and by replacing $\vecY$ with $\vecW$. 
Then, the update of $\bDelta_k$ is
\begin{equation*}
\ddot \bDelta_k= \frac{\text{diag}\left(\ddot\bGamma^{\top} \ddot \vecW_k \ddot\bGamma\right)}{\left|\text{diag}\left(\ddot\bGamma^{\top} \ddot \vecW_k \ddot\bGamma\right)\right|^\frac{1}{R}}
\label{eq:ve}
\end{equation*}

\item Model EV [$\matPsi_k=\bGamma_k\bDelta\bGamma_k^{\top}$] Here, the column covariance matrices have common shape but varying orientations.
By using the same approach of the EEV and VEV models, and by changing $\ddot\vecY$ with $\ddot\vecW$, we obtain the updates of $\bGamma_k$ and $\bDelta$ as
\begin{equation*}
\ddot\bGamma_k=\ddot \bL_k , \quad \ddot\bDelta = \frac{\sum\limits_{k=1}^K \ddot \vecOmega_k}{\left|\sum\limits_{k=1}^K \ddot \vecOmega_k\right|^\frac{1}{R}}. 
\label{eq:ev}
\end{equation*}

\item Model VV [$\matPsi_k=\bGamma_k\bDelta_k\bGamma_k^{\top}$] In the full unconstrained case, we obtain 
\begin{equation*}
\ddot \matPsi_k = \frac{\ddot \vecW_k}{\left|\ddot\vecW_k\right|^\frac{1}{R}}.
\label{eq:vv}
\end{equation*}

\end{itemize}

\subsubsection{A note on the initialization strategy}
\label{subsec:init}

To start our ECM algorithm, we followed the approach of \citet{tomarchio2020two}, where a generalization of the short-EM initialization strategy proposed by \citet{biernacki2003choosing} has been implemented.
It consists in $H$ short runs of the algorithm from several random positions.
The term ``short'' means that the algorithm is run for a few iterations $s$, without waiting for convergence.
In this manuscript, we set $H=100$ and $s=1$.
Then, the parameter set producing the largest log-likelihood is used to initialize the ECM algorithm.
In both simulated and real data analyses this procedure has shown stable results after multiple runs.

\section{Simulated analyses}
\label{sec:sim}

In this section, we examine different aspects of our HMMs through large-scale simulation studies.
Considering the high number of models, we will only focus on two of them for the sake of simplicity.
In detail, the EII-II HMM (which is the most parsimonious model) and the VVE-VE HMM (which is one of the two models for which an MM algorithm is used both for $\matsig_k$ and $\matPsi_k$) are considered.
For each model, several experimental conditions are evaluated.
Specifically, we set $P=R=2$, $I=100$, $T \in \left\{5,10,15\right\}$, $K\in\left\{2,4\right\}$ and two levels of overlap, that will be labeled as ``Overlap 1'' and ``Overlap 2''.
Therefore, $3 \times 2 \times 2 = 12$ scenarios are analyzed and, for each of them, 50 data sets are generated by the considered HMMs.

About the parameters used to generate the data, when $K=2$ we set 
\begin{itemize}
\item EII-II Model
\[
\matsig_1=\matsig_2=		
\begin{bmatrix}
    1.50 & 0.00 \\
    0.00 & 1.50 \end{bmatrix}, \quad
\matPsi_1=\matPsi_2=		
\begin{bmatrix}
    1.00 & 0.00 \\
    0.00 & 1.00 \end{bmatrix}, 
\]
\item VVE-VE Model
\[
\matsig_1=		
\begin{bmatrix}
    0.85 & 0.29 \\
    0.29 & 0.85 \end{bmatrix}, \
\matsig_2=
\begin{bmatrix}
    0.50 & 0.30 \\
    0.30 & 0.50 \end{bmatrix}, \ 
\matPsi_1=		
\begin{bmatrix}
    1.06 & 0.36 \\
    0.36 & 1.06 \end{bmatrix}, \
\matPsi_2=
\begin{bmatrix}
    1.25 & 0.75 \\
    0.75 & 1.25 \end{bmatrix},
\]
\end{itemize}
while for both HMMs we set $\bpi=\left(0.5,0.5\right)$, 
\[
\bPi= 
\begin{bmatrix}
    0.60 & 0.40 \\
    0.20 & 0.80 \end{bmatrix}, \quad		
\matm_1=
\begin{bmatrix}
    1.00 & 1.50 \\
    0.50 & 1.00 \end{bmatrix}. 
\]
The mean matrix of the second state ($\matm_2$) is obtained by adding a constant $c$ to each element of $\matm_1$, which depends on the level of overlap.
Specifically, we consider $c=2$ under the ``Overlap 1'' scenarios, whereas $c=5$ under the ``Overlap 2'' scenarios. 

When $K=4$, the first two hidden states have the same $\left\{\matsig_k,\matPsi_k,\matm_k; k=1,2\right\}$ as before.
Clearly, the covariance matrices of the third and fourth hidden states for the EII-II Model are still equal to those of the first two states. 
On the contrary, for the VVE-VE model we have 
\[
\matsig_3=		
\begin{bmatrix}
    1.45 & 1.05 \\
    1.05 & 1.45 \end{bmatrix}, \
\matsig_4=
\begin{bmatrix}
    1.33 & 0.29 \\
    0.29 & 1.33 \end{bmatrix}, \
\matPsi_3=		
\begin{bmatrix}
    1.45 & 1.00 \\
    1.00 & 1.45 \end{bmatrix}, \
\matPsi_4=
\begin{bmatrix}
    1.03 & 0.23 \\
    0.23 & 1.03 \end{bmatrix}.
\]
Then, for both HMMs we set $\bpi=\left(0.25,0.25,0.25,0.25\right)$ and
\[\bPi= 
\begin{bmatrix}
    0.55 & 0.00 & 0.21 & 0.24\\
    0.03 & 0.52 & 0.18 & 0.27\\   
    0.06 & 0.15 & 0.49 & 0.30 \\
		0.09 & 0.12 & 0.33 & 0.46 \end{bmatrix}.
\]
To obtain $\matm_3$ and $\matm_4$ we add $c=4$ and $c=-2$ to each element of $\matm_1$, respectively.

\subsection{Discussion}
\label{sec:pr}

First of all, we evaluate the recovery and the consistency of the estimated parameters by computing the mean square errors (MSEs).
Considering the high number of parameters that should be reported, we follow an approach similar to the one used by \citet{farcomeni2019robust}, i.e.~we calculate the average of the MSEs of each parameter of the model over the $K$ states, allowing us to summarize in a single number the MSE of each parameter.
Furthermore, before showing the obtained results, it is important to underline the well-known label switching issue, caused by the invariance of the likelihood function under relabeling the model states \citep{fruhwirth2006finite}.
There are no generally accepted labeling methods, and we simply attribute the labels by looking at the estimated $\matm_k$.

Table~\ref{tab:res1} and Table~\ref{tab:res2} report the average MSEs, computed after fitting the EII-II and VVE-VE HMMs, with the corresponding $K$, to the respective data sets.
Note that the $\matPsi$ covariance matrix is not reported in Table~\ref{tab:res1} since it is not estimated in the EII-II HMM.
\begin{table}[!htt]
    \centering
\caption{Average of the MSEs of the parameter estimates, over the $K$ states and 50 data sets, for the EII-II HMM under each scenario.}
\label{tab:res1}       % Give a unique label
       \begin{tabular}{ccccc|ccc}
\hline\noalign{\smallskip}
  $K$  &    Parameter       & \multicolumn{3}{c}{Overlap 1} & \multicolumn{3}{c}{Overlap 2} \\
\hline\noalign{\smallskip}
	  &           & $T=5$ & $T=10$ & $T=15$ & $T=5$ & $T=10$ & $T=15$ \\
 2  & $\matm$	  & 0.0083 & 0.0040 & 0.0024 & 0.0063 & 0.0033 & 0.0023 \\
    & $\matsig$ & 0.0020 & 0.0016 & 0.0007 & 0.0016 & 0.0012 & 0.0007 \\
		& $\bpi$	  & 0.0026 & 0.0029 & 0.0023 & 0.0024 & 0.0029 & 0.0023 \\
		& $\bPi$	  & 0.0013 & 0.0004 & 0.0004 & 0.0010 & 0.0005 & 0.0004 \\
\hline\noalign{\smallskip}
 4  & $\matm$	  & 0.0164 & 0.0084 & 0.0055 & 0.0135 & 0.0069 & 0.0044 \\
    & $\matsig$ & 0.0029 & 0.0010 & 0.0007 & 0.0024 & 0.0013 & 0.0010 \\
		& $\bpi$	  & 0.0022 & 0.0017 & 0.0022 & 0.0017 & 0.0021 & 0.0020 \\
    & $\bPi$	  & 0.0009 & 0.0006 & 0.0004 & 0.0009 & 0.0004 & 0.0003 \\
\noalign{\smallskip}\hline
        \end{tabular}% <------ Don't forget this %
\end{table}
\begin{table}[!htt]
    \centering
\caption{Average of the MSEs of the parameter estimates, over the $K$ states and 50 data sets, for the VVE-VE HMM under each scenario.}
\label{tab:res2}       % Give a unique label
       \begin{tabular}{ccccc|ccc}
\hline\noalign{\smallskip}
  $K$  &    Parameter       & \multicolumn{3}{c}{Overlap 1} & \multicolumn{3}{c}{Overlap 2} \\
\hline\noalign{\smallskip}
	  &           & $T=5$ & $T=10$ & $T=15$ & $T=5$ & $T=10$ & $T=15$ \\
 2  & $\matm$	  & 0.0052 & 0.0024 & 0.0015 & 0.0033 & 0.0017 & 0.0012 \\
    & $\matsig$ & 0.0021 & 0.0010 & 0.0006 & 0.0014 & 0.0008 & 0.0005 \\
    & $\matPsi$ & 0.0028 & 0.0014 & 0.0009 & 0.0020 & 0.0008 & 0.0007 \\
		& $\bpi$	  & 0.0042 & 0.0032 & 0.0030 & 0.0029 & 0.0030 & 0.0020 \\
    & $\bPi$	  & 0.0010 & 0.0007 & 0.0003 & 0.0009 & 0.0005 & 0.0004 \\
\hline\noalign{\smallskip}
 4  & $\matm$	  & 0.0176 & 0.0092 & 0.0050 & 0.0094 & 0.0050 & 0.0035 \\
    & $\matsig$ & 0.0110 & 0.0049 & 0.0032 & 0.0058 & 0.0030 & 0.0022 \\
    & $\matPsi$ & 0.0076 & 0.0038 & 0.0023 & 0.0041 & 0.0026 & 0.0016 \\
		& $\bpi$	  & 0.0030 & 0.0021 & 0.0024 & 0.0020 & 0.0024 & 0.0021 \\
    & $\bPi$	  & 0.0011 & 0.0006 & 0.0005 & 0.0007 & 0.0005 & 0.0003 \\
\noalign{\smallskip}\hline
        \end{tabular}% <------ Don't forget this %
\end{table}
As we can see, the MSEs can be considered negligible in all the considered scenarios.
It is interesting to note that, for a fixed overlap, their values become better with the increase of $T$ and that, fixed $T$, their values roughly improve as we move from ``Overlap 1'' to ``Overlap 2'', thus indicating a decrease in the level of overlap.
Additionally, when the VVE-VE HMM is considered, it seems that the MM algorithms used for estimating the covariance matrices produce reliable values.

\begin{figure}[!ht]
\subfloat[\label{fig:eii_c_2}]{
	\includegraphics[width=0.495\textwidth]{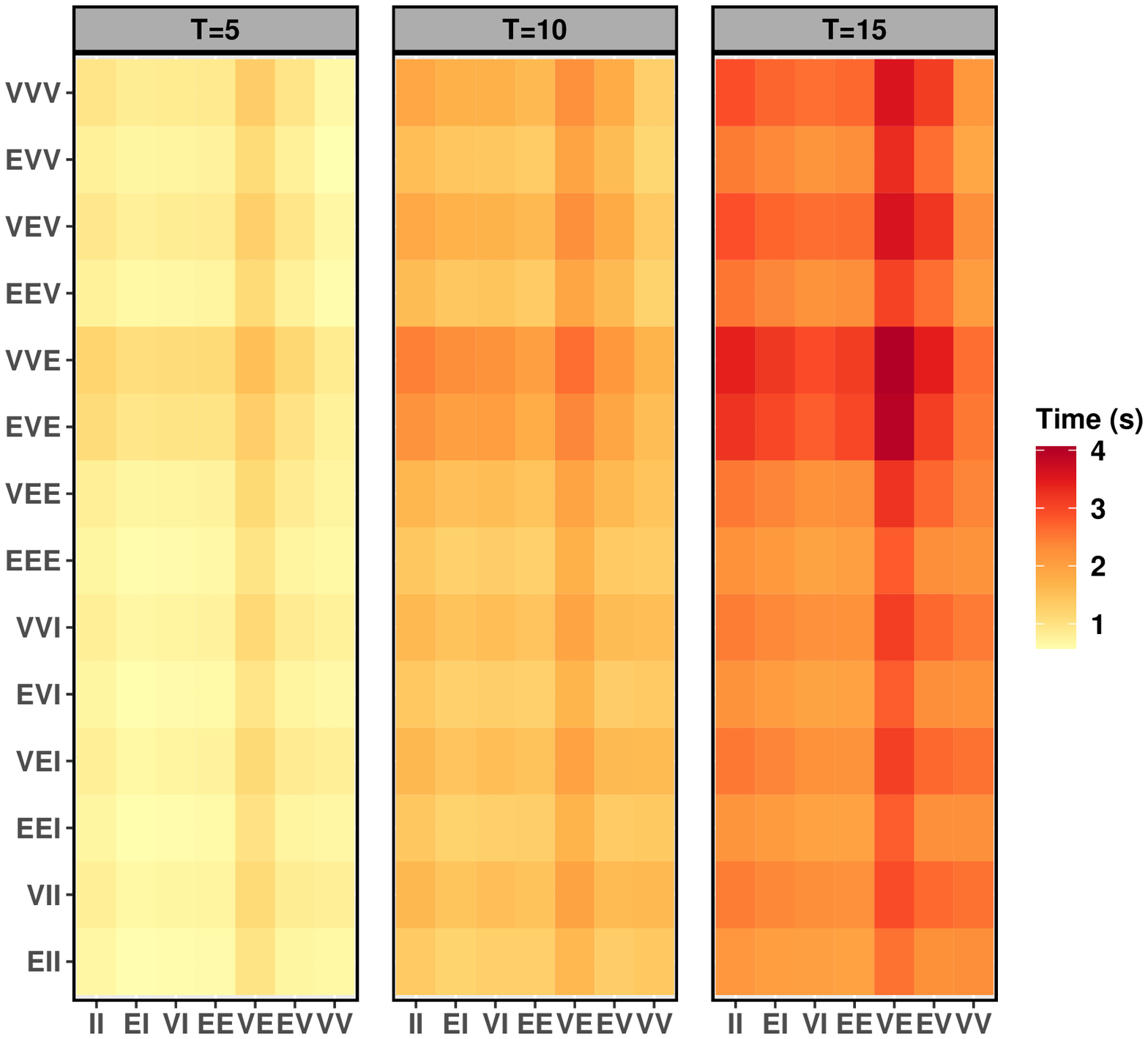}}
\subfloat[\label{fig:eii_f_2}]{%
	\includegraphics[width=0.495\textwidth]{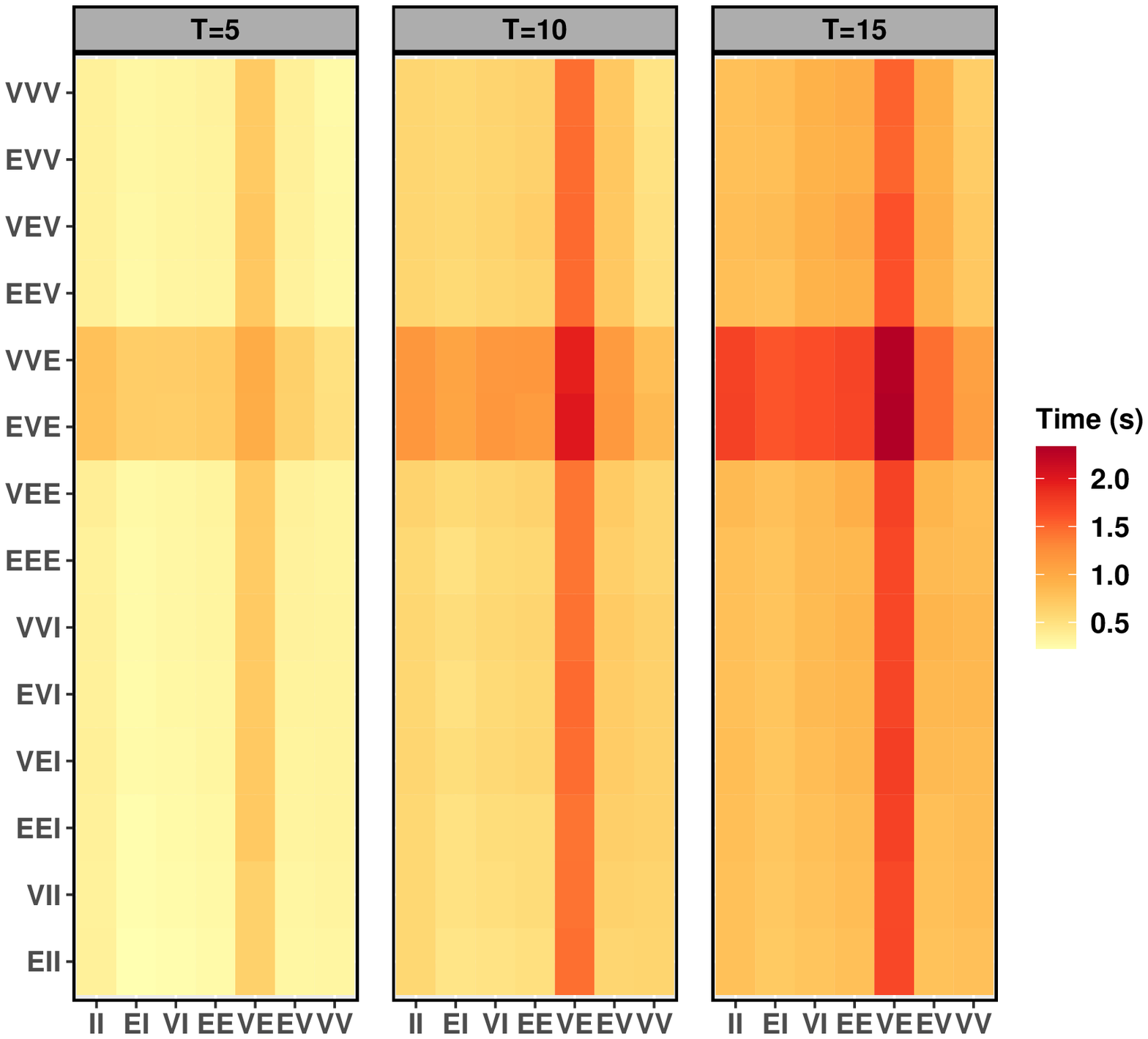}}
\caption{Heat maps of the average computational time for the 98 HMMs, computed over 50 data sets, when the data are generated by a EII-II HMM with $K=2$ and ``Overlap 1'' (a) or ``Overlap 2'' (b).}
\label{fig:1}       
\end{figure}

\begin{figure}[!ht]
\subfloat[\label{fig:eii_c_4}]{
  \includegraphics[width=0.495\textwidth]{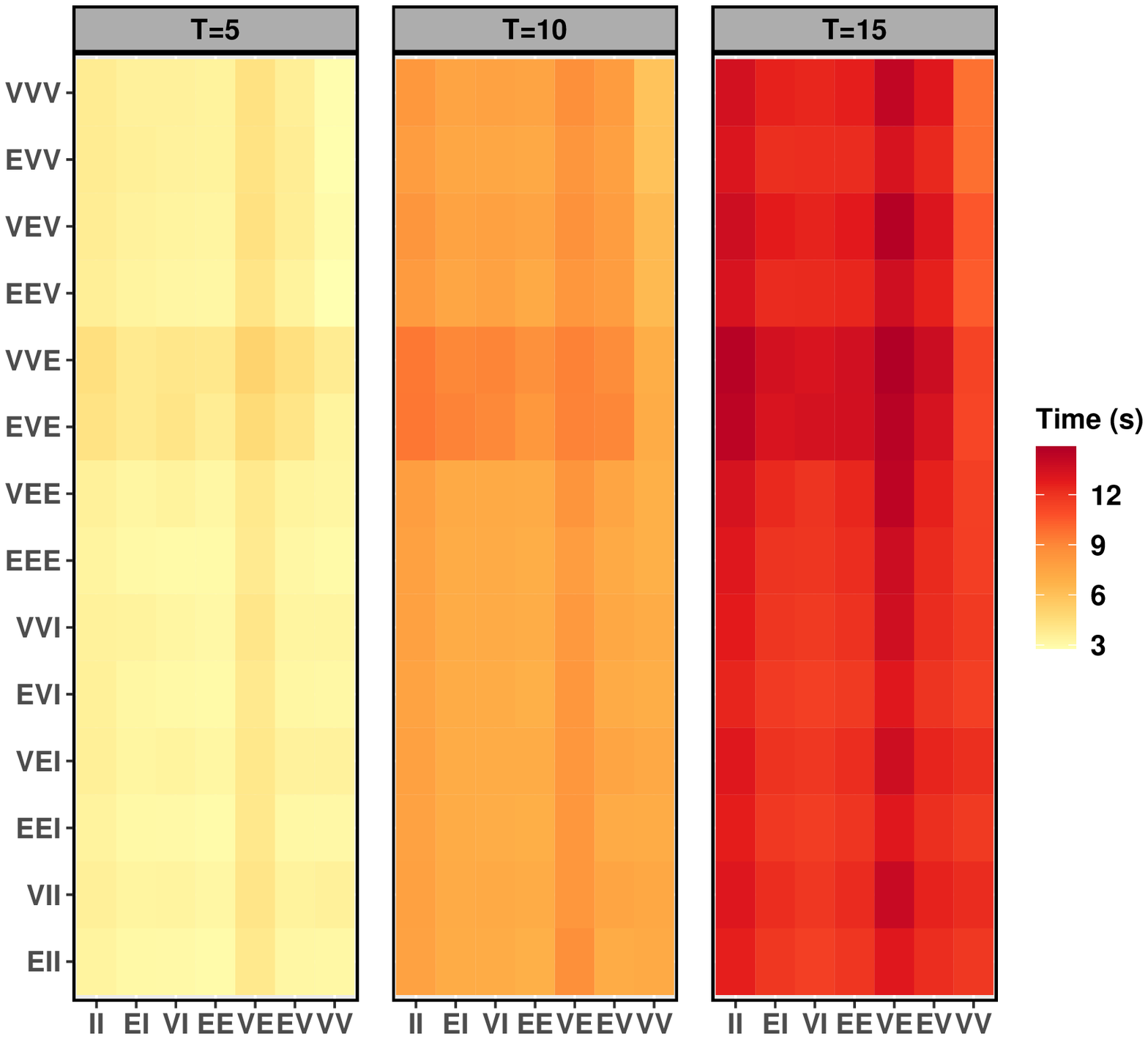}}
	\subfloat[\label{fig:eii_f_4}]{%
	\includegraphics[width=0.495\textwidth]{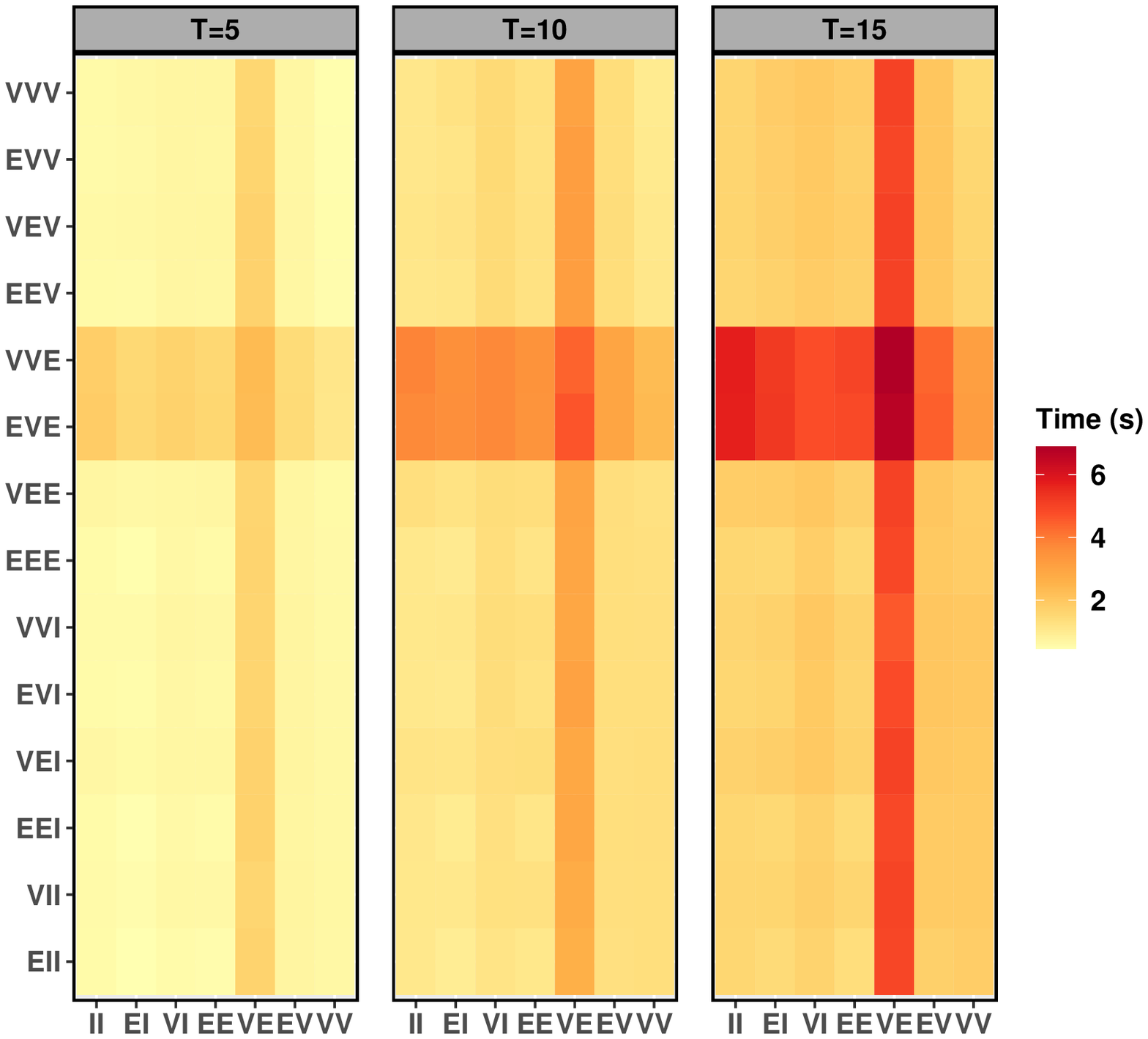}}
\caption{Heat maps of the average computational time for the 98 HMMs, computed over 50 data sets, when the data are generated by a EII-II HMM with $K=4$ and ``Overlap 1'' (a) or ``Overlap 2'' (b).}
\label{fig:2}       
\end{figure}

\begin{figure}[!ht]
\subfloat[\label{fig:vve_c_2}]{
  \includegraphics[width=0.495\textwidth]{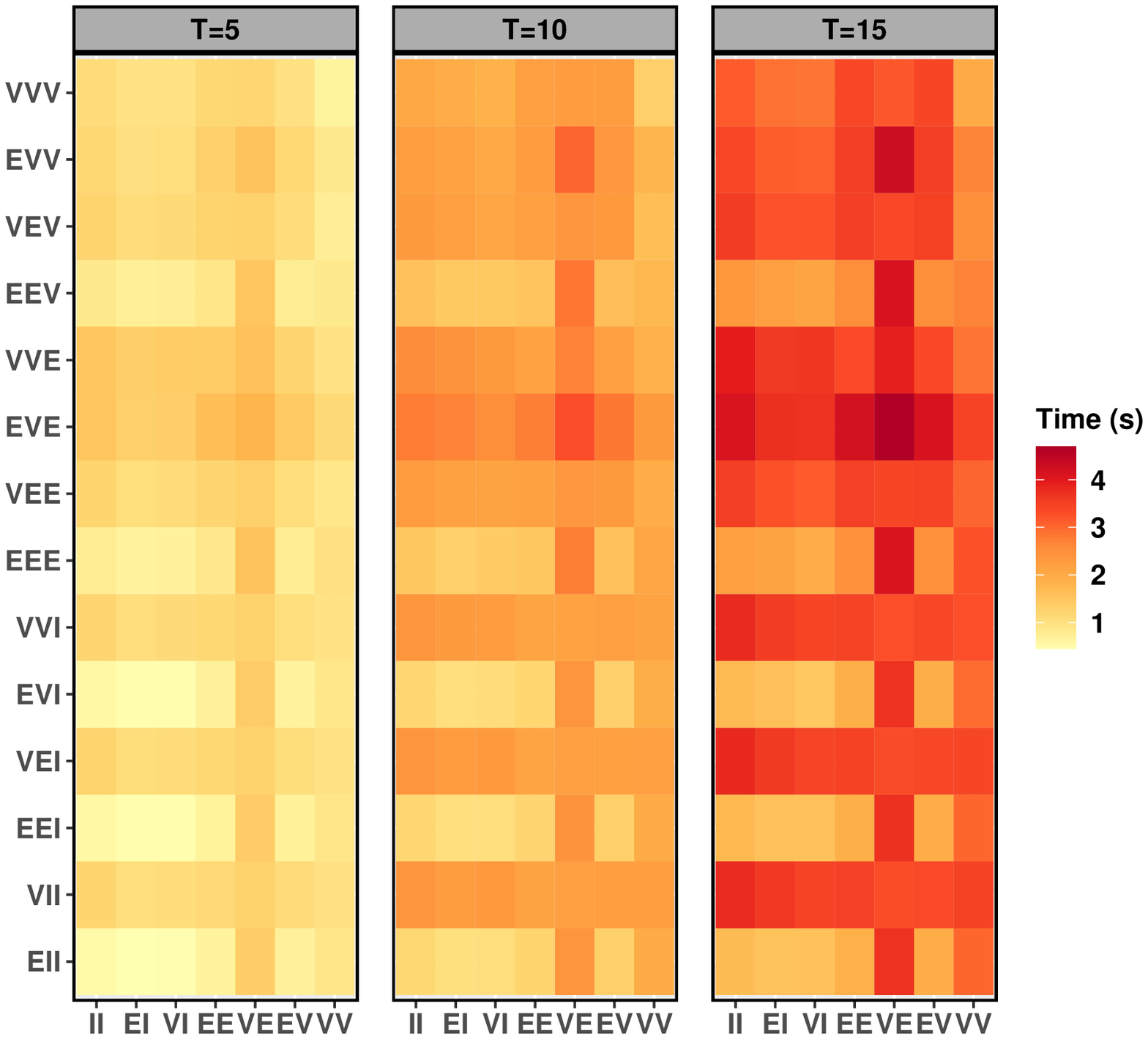}}
	\subfloat[\label{fig:vve_f_2}]{%
	\includegraphics[width=0.495\textwidth]{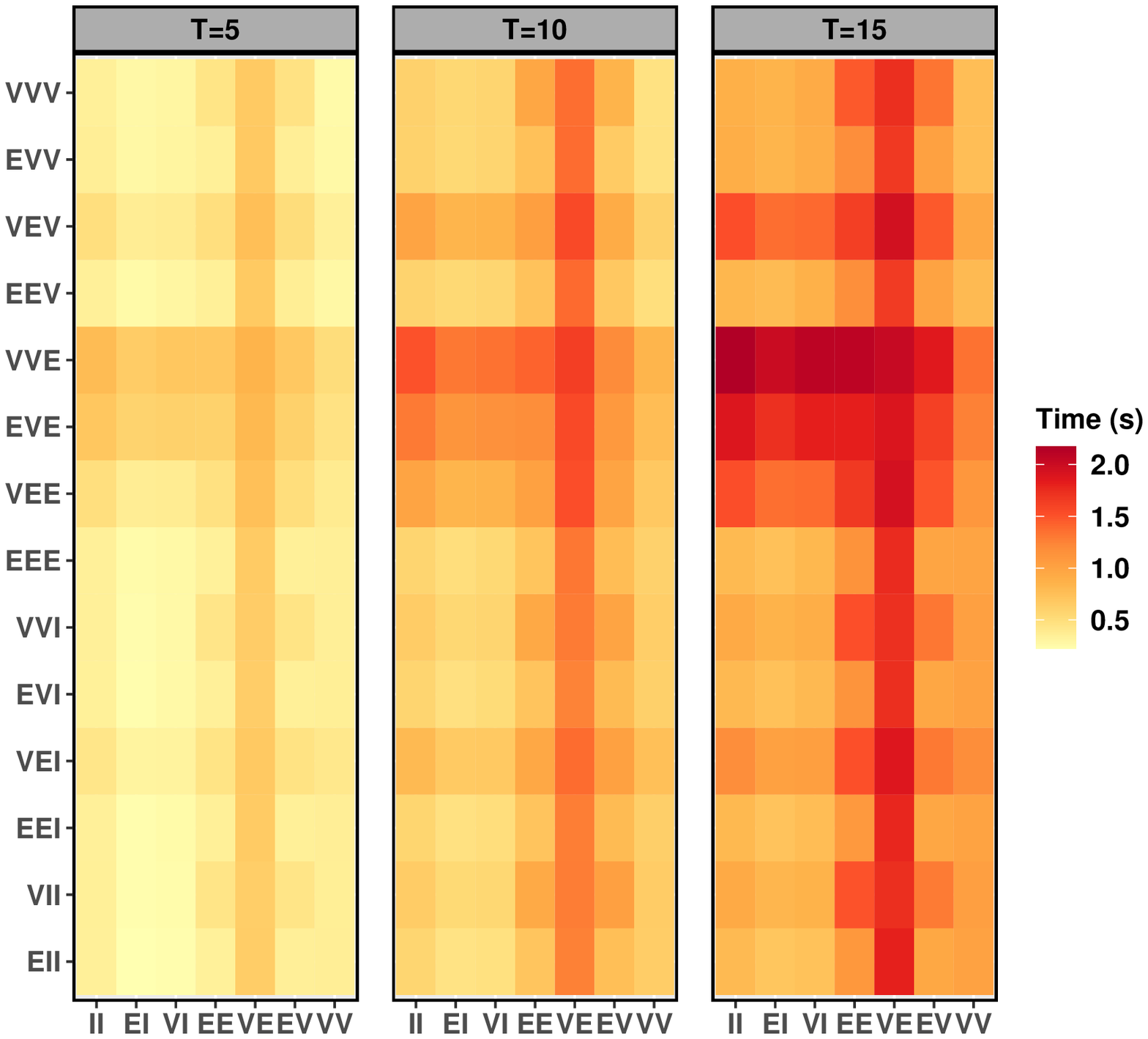}}
\caption{Heat maps of the average computational time for the 98 HMMs, computed over 50 data sets, when the data are generated by a VVE-VE HMM with $K=2$ and ``Overlap 1'' (a) or ``Overlap 2'' (b).}
\label{fig:3}      
\end{figure}

\begin{figure}[!ht]
\subfloat[\label{fig:vve_c_4}]{
  \includegraphics[width=0.495\textwidth]{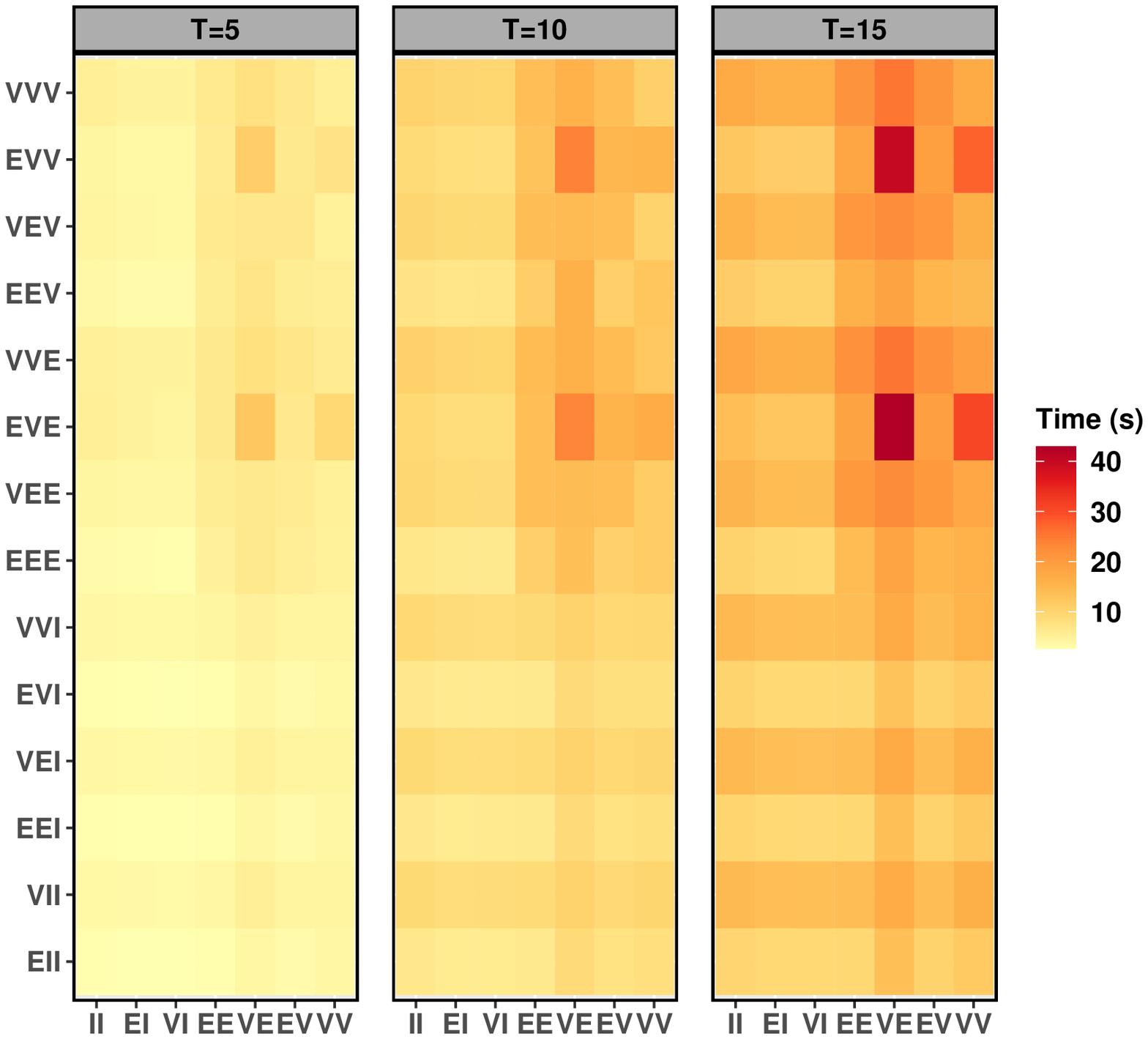}}
		\subfloat[\label{fig:vve_f_4}]{%
	\includegraphics[width=0.495\textwidth]{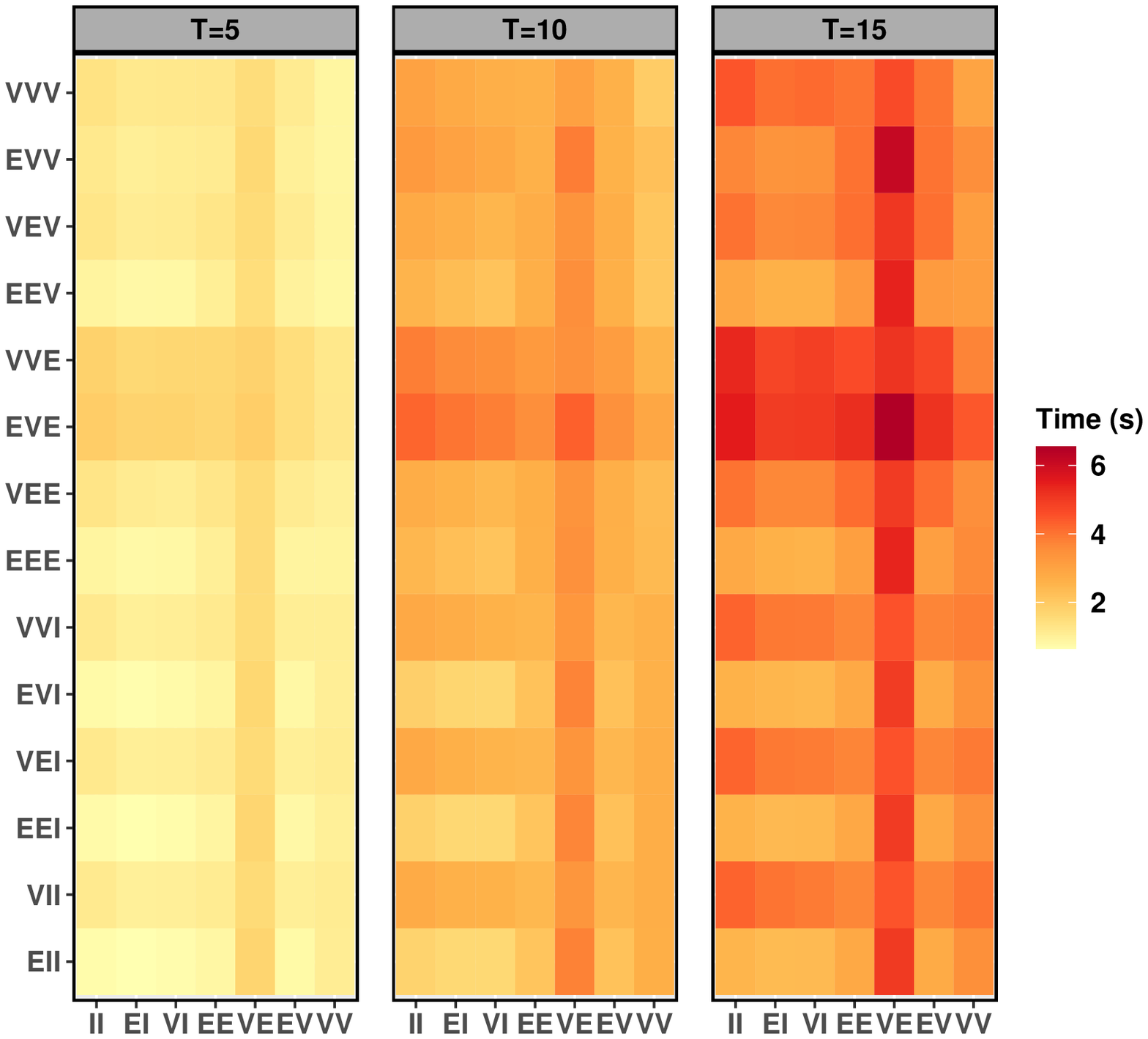}}
\caption{Heat maps of the average computational time for the 98 HMMs, computed over 50 data sets, when the data are generated by a VVE-VE HMM with $K=4$ and ``Overlap 1'' (a) or ``Overlap 2'' (b).}
\label{fig:4}       
\end{figure}

Another aspect that is interesting to evaluate, is the computational time required for fitting the HMMs.
In detail, on each of the above data sets, all the 98 HMMs are now fitted for the corresponding $K$, and their computational times (in seconds) are illustrated by using the heat maps of \Cref{fig:1,fig:2,fig:3,fig:4}.
Computation is performed on a Windows 10 PC, with AMD Ryzen 7 3700x CPU, 16.0 GB RAM, using the \textsf{R} 64-bit statistical software \citep{R}, and the \texttt{proc.time()} function of the \textbf{base} package is used to measure the time.
As it is reasonable to expect, the computational time grows as $T$ increases on each scenario, and it approximately halves when we pass from ``Overlap 1'' to ``Overlap 2'', highlighting the easier of estimation in this case.
Furthermore, with the exclusion of \figurename~\ref{fig:vve_c_4}, the computational time approximately triplicates when we move from fitting HMMs with $K=2$ to HMMs with $K=4$ hidden states.
It is interesting to note that the EVE-VE and VVE-VE HMMs, which are the two models for which we use a MM algorithm for estimating both covariance matrices, are the most time consuming, with a computational burden that seems to double with respect to the other models.
This is particularly evident in the ``Overlap 2'' scenarios.

The total computational time can be strongly reduced by exploiting parallel computing.
In detail, Table~\ref{tab:res4} shows the overall time taken by fitting the 98 HMMs sequentially (default in \textsf{R}) and by parallelizing them on 14 cores.
As we can see, the computational burden is decreased by about 10 times, and all the models can be fitted in a reasonable fast way (with some exceptions in the ``Overlap 1'' scenarios).
\begin{table}[!htt]
    \centering
\caption{Computational times (in seconds) for running the algorithm sequentially or via parallel computing. They refer to the fit of all the 98 HMMs with $K$ states, averaged over the 50 data sets, and generated by the two HMMs on each scenario.}
\label{tab:res4}       % Give a unique label
       \begin{tabular}{cccccc|ccc}
\hline\noalign{\smallskip}
	Type  & HMM  &  $K$  & \multicolumn{3}{c}{Overlap 1} & \multicolumn{3}{c}{Overlap 2} \\
\hline\noalign{\smallskip}
	  &   &       & $T=5$ & $T=10$ & $T=15$ & $T=5$ & $T=10$ & $T=15$ \\
 \multirow{3}{*}{Sequential} &  \multirow{2}{*}{EII-II}  & 2   & 82.38  & 159.96 & 252.27 & 41.37 & 78.03 & 103.66 \\
                             &                           & 4   & 346.51 & 753.71 & 1236.52& 89.32 & 176.54 & 259.24 \\
                             &  \multirow{2}{*}{VVE-VE}  & 2   & 103.31 & 199.63 & 301.53 & 43.19 & 84.66 & 123.98 \\
                             &                           & 4   & 473.74 & 1018.19 & 1551.68 & 116.69 & 270.18 & 380.48 \\
\hline\noalign{\smallskip}
 \multirow{3}{*}{Parallel  } &  \multirow{2}{*}{EII-II}  & 2   & 9.65 & 15.53 & 22.68 & 6.63 & 9.42 & 11.33 \\
                             &                           & 4   & 29.87 & 60.01 & 95.54 & 10.39 & 17.34 & 23.82 \\
                             &  \multirow{2}{*}{VVE-VE}  & 2   & 11.12 & 18.43 & 26.06 & 6.70 & 9.75 & 12.79 \\
                             &                           & 4   & 43.11 & 86.26 & 130.60 & 12.19 & 23.77 & 32.46 \\
\noalign{\smallskip}\hline
        \end{tabular}% <------ Don't forget this %
\end{table}

Lastly, the capability of the Bayesian information criterion (BIC; \citealp{schwarz1978estimating}) in identifying the true parsimonious structure and the correct
number of groups is investigated.
This is because, so far, we have fitted models with $K$ equal to the true number of states present in the data, and we need to assess if the BIC, which is one of the most famous and used tools in model-based clustering, accurately works.
Therefore, on each of the above data sets, the 98 HMMs are fitted for $K\in\left\{1,\ldots,K+1\right\}$, and the results are reported in Table~\ref{tab:res3}.
First of all, in each scenario, the true $K$ has been always selected by the best fitting model according to the BIC (for this reason this information is not reported in Table~\ref{tab:res3}).
Additionally, we notice that in almost all the cases the true data generating model has been identified by the BIC.
In those few cases where the BIC selects a wrong model, this is because of an incorrect choice of the parsimonious structure for one of the two covariance matrices $\matsig$ or $\matPsi$.
\begin{table}[!htt]
    \centering
\caption{Number of times, over the 50 data sets generated by the two HMMs on each scenario, for which the true parsimonious structure is selected by the BIC when all the 98 HMMs are fitted for $k\in\left\{1,\ldots,K+1\right\}$.}
\label{tab:res3}       % Give a unique label
       \begin{tabular}{ccccc|ccc}
				\hline\noalign{\smallskip}
HMM &  $K$  & \multicolumn{3}{c}{Overlap 1} & \multicolumn{3}{c}{Overlap 2} \\
\hline\noalign{\smallskip}
	  &          & $T=5$ & $T=10$ & $T=15$ & $T=5$ & $T=10$ & $T=15$ \\
 EII-II  & 2   & 47 & 48 & 48 & 49 & 49 & 47 \\
         & 4   & 46 & 50 & 50 & 50 & 50 & 49 \\
\hline\noalign{\smallskip}
 VVE-VE  & 2	 & 45 & 49 & 48 & 48 & 49 & 48 \\
         & 4   & 47 & 49 & 50 & 50 & 50 & 50 \\
\noalign{\smallskip}\hline
        \end{tabular}% <------ Don't forget this %
\end{table}

\section{Real data example}
\label{sec:real}

In this section, we analyze data concerning the unemployment rate in the Italian provinces (NUTS3, according to the European Nomenclature of Territorial Units for Statistics).
The data comes from the Italian National Institute of Statistics (ISTAT), a public research organization and the main producer of official statistics in the service of citizens and policy-makers in Italy, and are freely accessible at \url{http://dati.istat.it/#}.
In detail, we investigate the $I=98$ Italian provinces for which the unemployment rate is available from the beginning of the data collection at the provincial level (2004) to the most recent year (2019).
This implies that we are considering $T=16$ years of data.
Note that some provinces are not included in the analysis since the data were available for only few years.

For each province, the unemployment rate is recorded in a two-factor format.
The first factor, gender, has two levels (i.e.~$P=2$): males and females. 
The second factor, age, has three levels (i.e.~$R=3$) driven by the age category: 15--24, 25--34 and 35--older.
Therefore, the whole data set is presented in a four-way array having dimensions $2 \times 3 \times 98 \times 16$.
The unemployment rates are then mapped to the real line by using the logit transformation, as commonly done in this branch of literature for overcoming boundary bias problems (see, e.g. \citealp{wallis1987time,koop1999dynamic,hudomiet2015role}).

In analyzing this data set, several questions arise concerning the existence of areas with similar unemployment levels among the Italian provinces.
Historically Southern Italy have always shown worse economic performance with respect to the rest of the Country \citep{daniele2007prodotto}.  
However, within each region (NUTS2, according to the European Nomenclature of Territorial Units for Statistics) there can be considerable differences among the provinces that are part of it.
Also of interest is the strength of time dependence as measured by the transition probability matrix, as well as how the provinces move between the hidden states.
This latter aspect can be particularly of interest in light of the two main recessions that the Italian economy faced in 2008 and 2011. 
Relatedly, an overview of which provinces have best withstood the crises or which have been able to recover from the crises can be easily obtained.

\subsection{Discussion}
\label{sec:pr2}

Our 98 HMMs are fitted to the data for $K\in\left\{1,\ldots,10\right\}$ and, according to the BIC, the best model is the VEV-EE with $K=7$ hidden states.
Despite the logit scale, we can easily interpret the $K=7$ states by looking at the estimated mean matrices
\[
\matm_1=
\begin{bmatrix}
    -2.34 & -3.50 & -4.14 \\
    -1.84 & -2.82 & -3.33 \end{bmatrix}, \quad
\matm_2=
\begin{bmatrix}
    -1.61 & -2.88 & -3.57 \\
    -1.27 & -2.39 & -3.05 \end{bmatrix}, 
\]
\[
\matm_3=
\begin{bmatrix}
    -1.67 & -2.87 & -3.38\\
    -1.27 & -2.08 & -2.69 \end{bmatrix}, \quad
\matm_4=
\begin{bmatrix}
    -1.05 & -2.33 & -3.02 \\
    -0.80 & -1.93 & -2.68 \end{bmatrix}, 
\]
\[
\matm_5=
\begin{bmatrix}
    -0.80 & -1.86 & -2.83\\
    -0.44 & -1.36 & -2.35 \end{bmatrix}, \quad
\matm_6=
\begin{bmatrix}
    -0.37 & -1.38 & -2.31 \\
    -0.09 & -1.03 & -2.10 \end{bmatrix}, 
\]
\[
\matm_7=
\begin{bmatrix}
    0.14 & -0.85 & -1.81\\
    0.39 & -0.53 & -1.70 \end{bmatrix}.
\]
As we can note, it is possible to sort the states according to growing unemployment levels, both in the gender and ages factors.
Specifically, as we move from the first to the seventh state the unemployment rises, and each state becomes worse than the previous ones under each point of view.
The only occasion where this happens partially concerns the states two and three.
Indeed, regardless of gender, state two shows better rates for people over 25 and conversely state three is preferred for people under 25.
However, since 4 times out of 6 state two is preferred, and considering that this involves the majority of people, we could globally consider it better than state three. 
We can also observe that, regardless of the considered state, the unemployment levels are higher for females and get lower as the age increases.
It might be also interesting to report that, regardless of the age class, the lowest relative differences between the two genders are in the seventh state, the worst.

Useful insights can also be obtained from the analysis of the estimated covariance matrices.
The best fitting model suggests various volumes and orientations but equal shapes for the gender-related covariance matrices, whereas all shapes and orientations are found to be the same for the age-based covariance matrices.
In detail, the estimated gender-related covariance matrices are
\[
\matsig_1=
\begin{bmatrix}
     0.27 & 0.03 \\
     0.03 & 0.20  \end{bmatrix}, \hspace{0.6mm}
\matsig_2=
\begin{bmatrix}
     0.13 & 0.01 \\
     0.01 & 0.10   \end{bmatrix}, \hspace{0.6mm}
\matsig_3=
\begin{bmatrix}
    0.36 & 0.02 \\
    0.02 & 0.25  \end{bmatrix}, 
\]
\[
\matsig_4=
\begin{bmatrix}
    0.11 & 0.02 \\
    0.02 & 0.10 \end{bmatrix}, \hspace{0.6mm}
\matsig_5=
\begin{bmatrix}
    0.09 & 0.02 \\
    0.02 & 0.12  \end{bmatrix}, \hspace{0.6mm}
\matsig_6=
\begin{bmatrix}
    0.12 & 0.02 \\
    0.02 & 0.16 \end{bmatrix}, \hspace{0.6mm}
\matsig_7=
\begin{bmatrix}
    0.09 & 0.01 \\
    0.01 & 0.11 \end{bmatrix}.
\]
We notice that, excluding the first and third, all the states have similar covariance matrices.
Furthermore, while the first four states have higher variances for men, the last three show greater variances for women.
Additionally, it is interesting to see that the third state has the largest variances both for both men and women.
With respect to the age-based covariance matrices, showed in following,
\[
\matPsi_1,\ldots,\matPsi_7=
\begin{bmatrix}
    1.66 & 0.18 & 0.19 \\
    0.18 & 1.02 & 0.11 \\
		0.19 & 0.11 & 0.63 \end{bmatrix}, 
\]
we can note that the variances become lower as the age categories grows. 

\begin{figure}[!htt]
\centering
\includegraphics[width=0.5\textwidth]{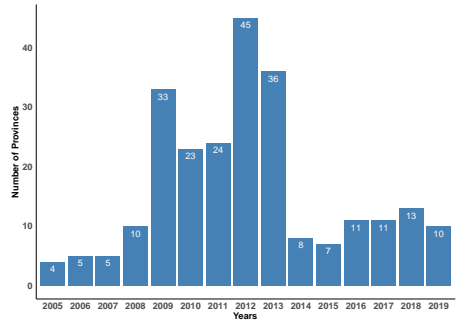}
\caption{Number of provinces that changed state in each year (from the second year onwards).} 
\label{fig:5}       
\end{figure}

Lastly, before showing how these states cluster the Italian provinces, it is worth analyzing the estimated transition probability matrix 
\[
\bPi=
\begin{bmatrix}
    0.80 & 0.20 & 0.00 & 0.00 & 0.00 & 0.00 & 0.00 \\
    0.03 & 0.72 & 0.01 & 0.24 & 0.00 & 0.00 & 0.00 \\
		0.00 & 0.14 & 0.72 & 0.05 & 0.06 & 0.03 & 0.00 \\ 
		0.00 & 0.03 & 0.08 & 0.83 & 0.06 & 0.00 & 0.00 \\ 
		0.00 & 0.00 & 0.00 & 0.02 & 0.84 & 0.14 & 0.00 \\
		0.00 & 0.00 & 0.01 & 0.00 & 0.04 & 0.86 & 0.09 \\
		0.00 & 0.00 & 0.00 & 0.00 & 0.00 & 0.04 & 0.96 \end{bmatrix},
\]
and how many provinces have changed their state in each year (from the second year onwards).
The latter aspect is illustrated in the bar plot of Figure~\ref{fig:5}.
As we can note by the estimated transition probability matrix, transitions between states are not uncommon between adjacent states, whereas are null among distant states.
Furthermore, it seems that the persistence of staying in a state, increases as we move from the fourth to the seventh, i.e.~it appears increasingly difficult for the provinces clustered in the troubled states to improve their position.
In particular, the last state shows a high persistence.

From the analysis of the bar plot, the highest number of switches between the states occurs over the years 2009--2013, as a consequence of the two aforementioned economic recessions.
Indeed, all the provinces showed an increase in the unemployment rates in those years, causing a change towards worse states.
This can be better understood by looking at the Italian provinces maps of \Cref{fig:6,fig:7}, that are colored according to state memberships.
Note that the provinces not included in the analysis are colored in gray.
For simplicity, we avoid to plot a map for each of the 16-years of data, and we limit to report some key years.
It is interesting to note that, in almost all the cases, not all the provinces belonging to the same region are clustered in the same state.
Therefore, working at province-level may offer additional insight than working with regional-level.
\begin{figure}[!htt]
\subfloat[\label{fig:2004}]{
	\includegraphics[width=0.44\textwidth]{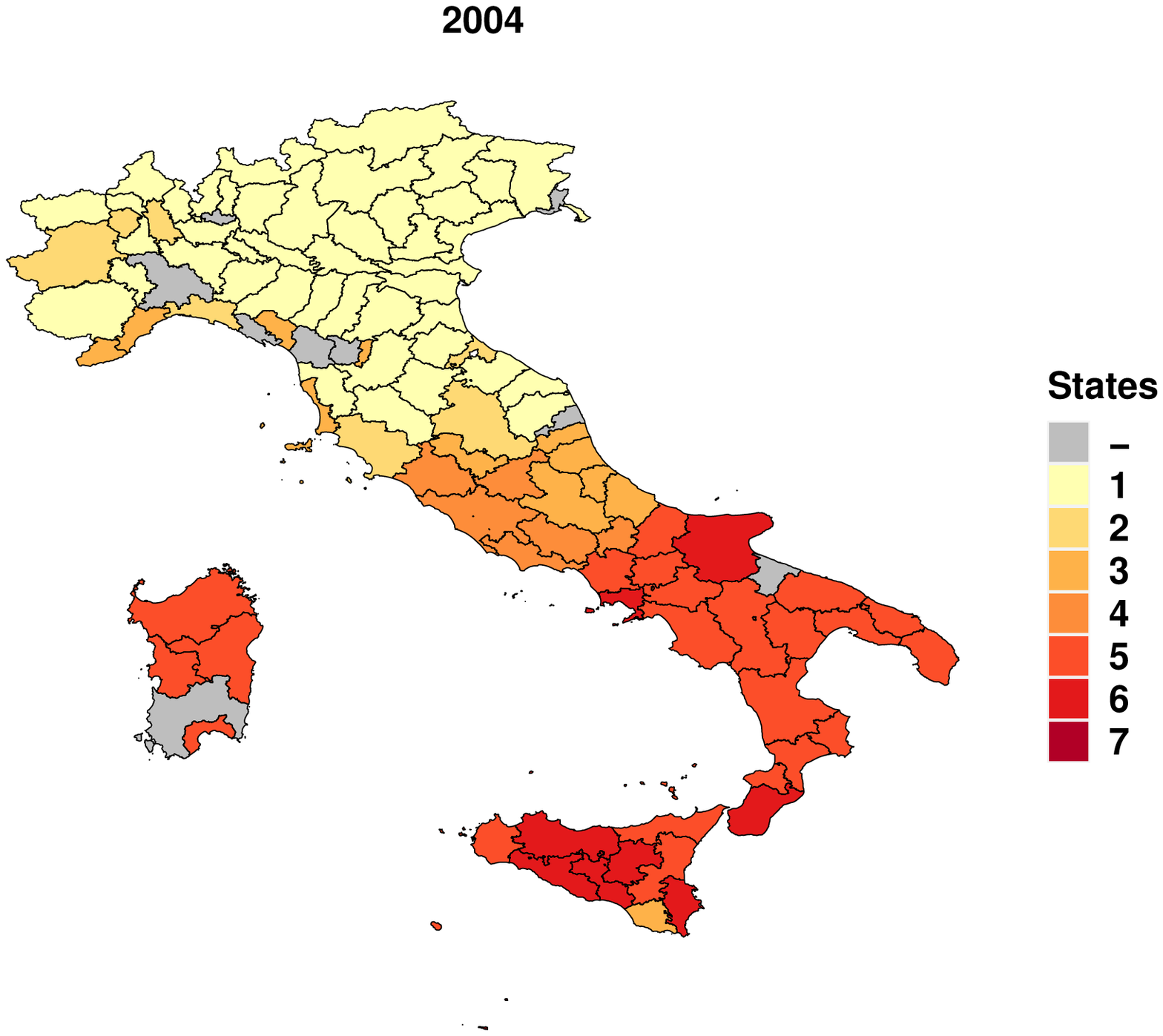}}\hspace{0.5em}
\subfloat[\label{fig:2009}]{%
	\includegraphics[width=0.44\textwidth]{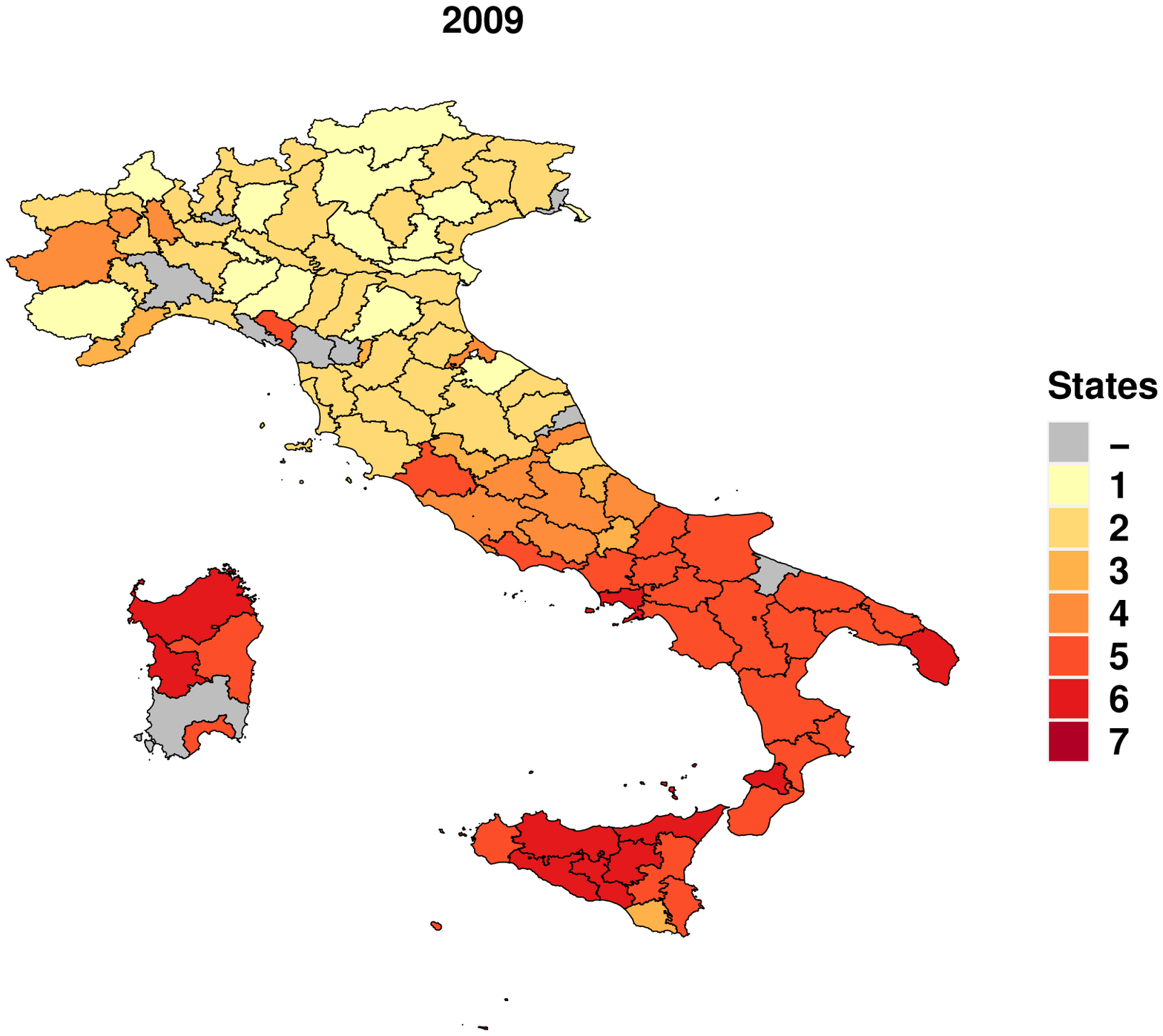}}
\\
\subfloat[\label{fig:2010}]{
	\includegraphics[width=0.44\textwidth]{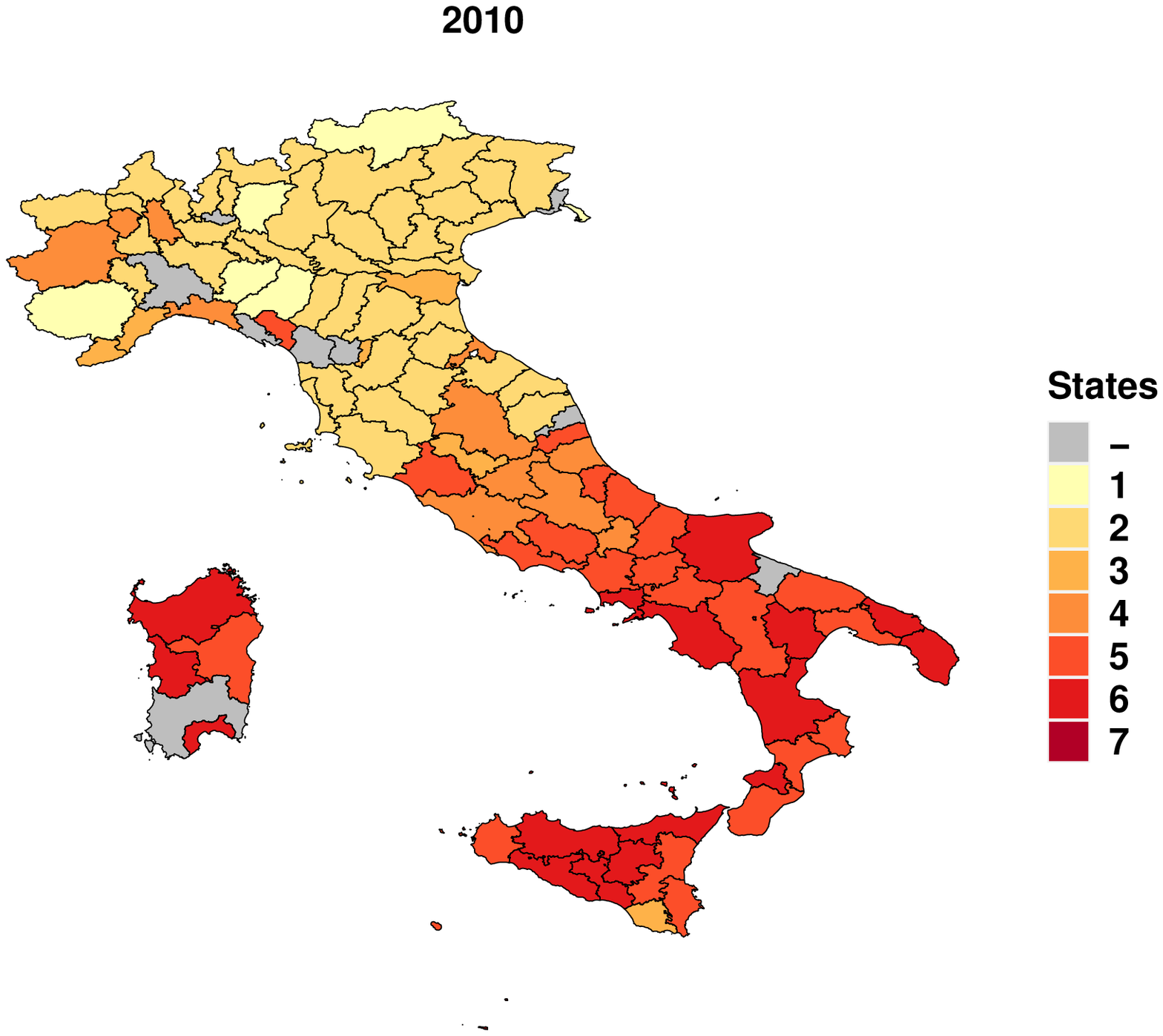}}\hspace{0.5em}
\subfloat[\label{fig:2011}]{%
	\includegraphics[width=0.44\textwidth]{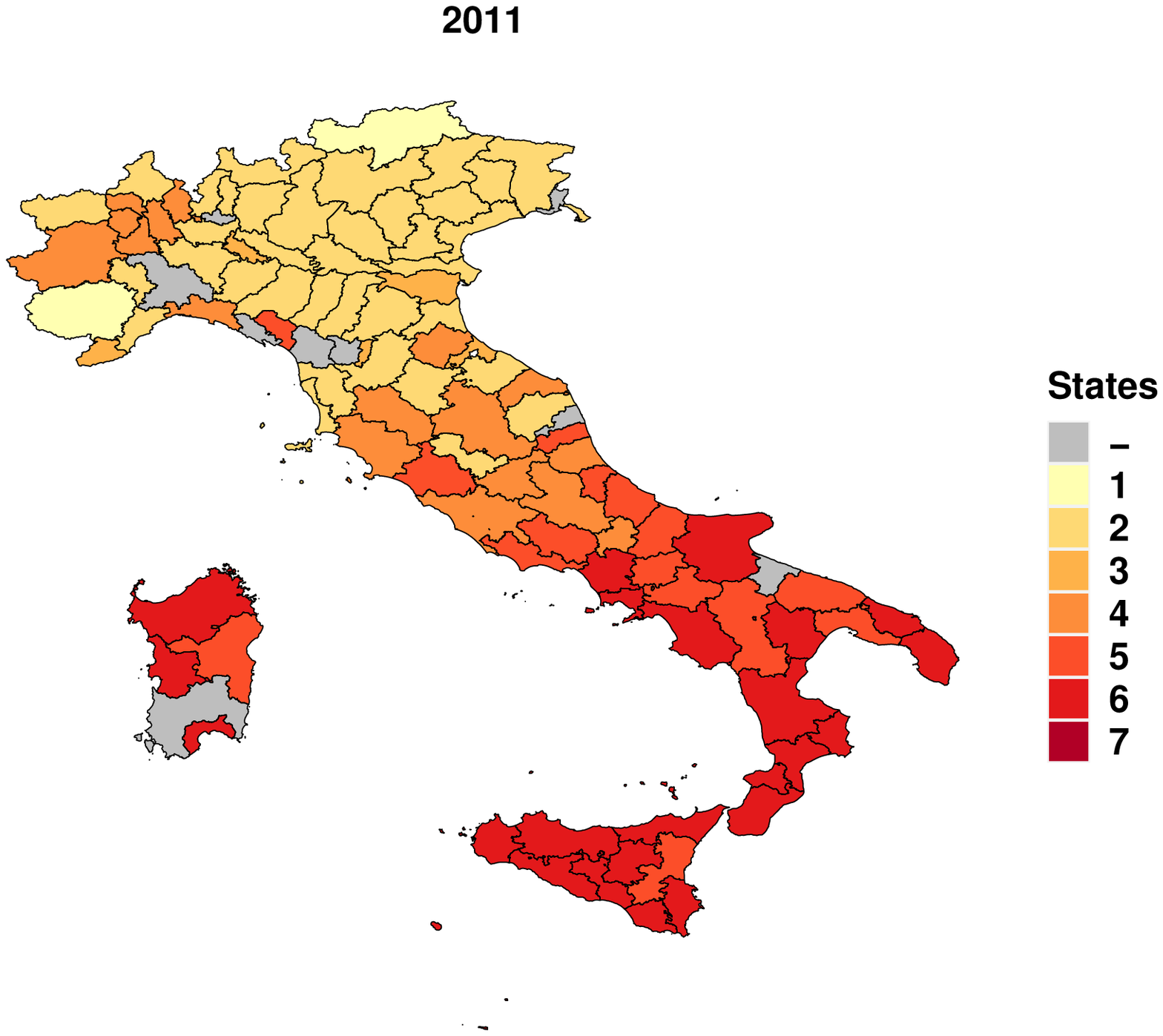}}
\\
\subfloat[\label{fig:2012}]{
	\includegraphics[width=0.44\textwidth]{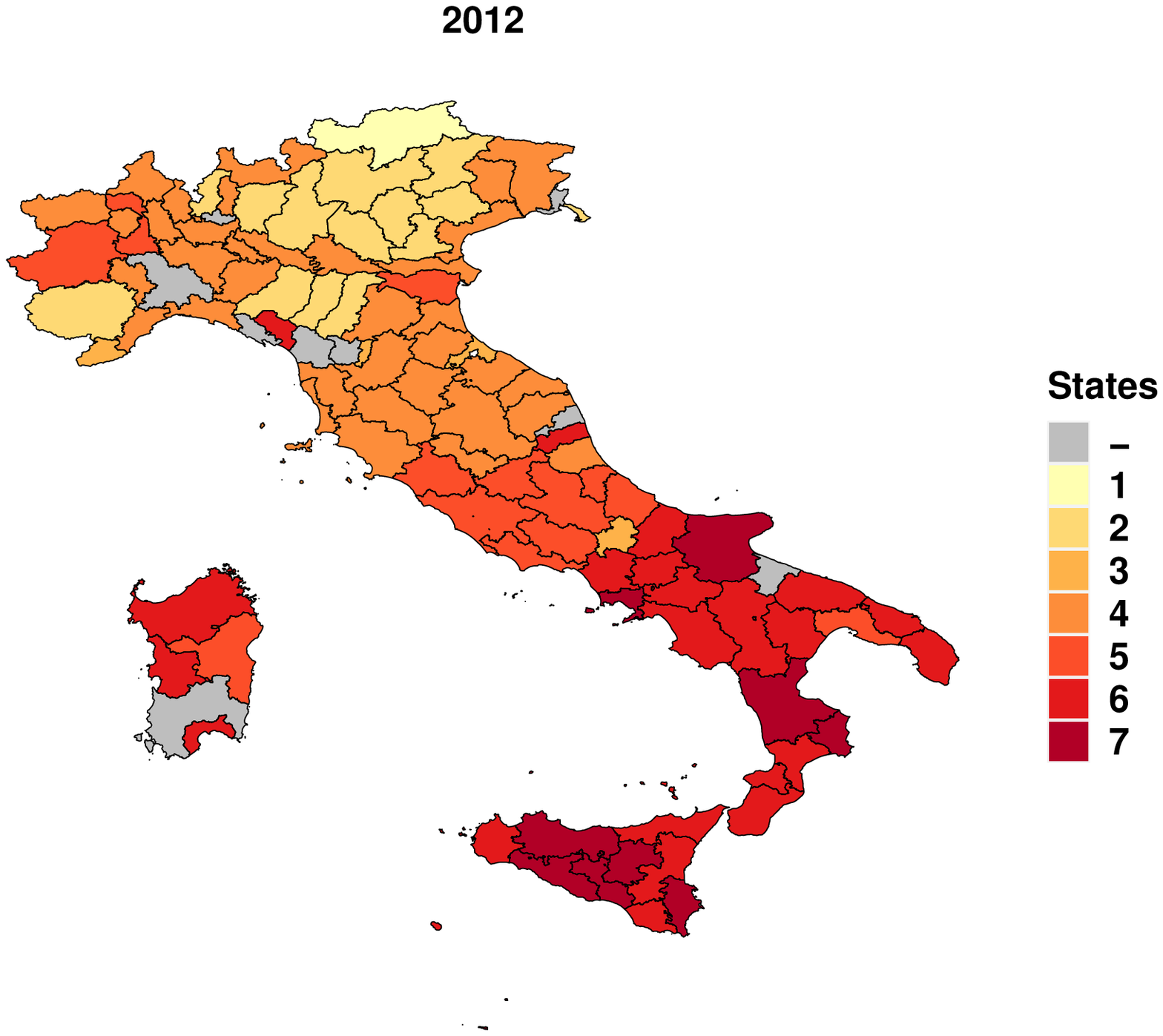}}\hspace{0.5em}
\subfloat[\label{fig:2013}]{%
	\includegraphics[width=0.44\textwidth]{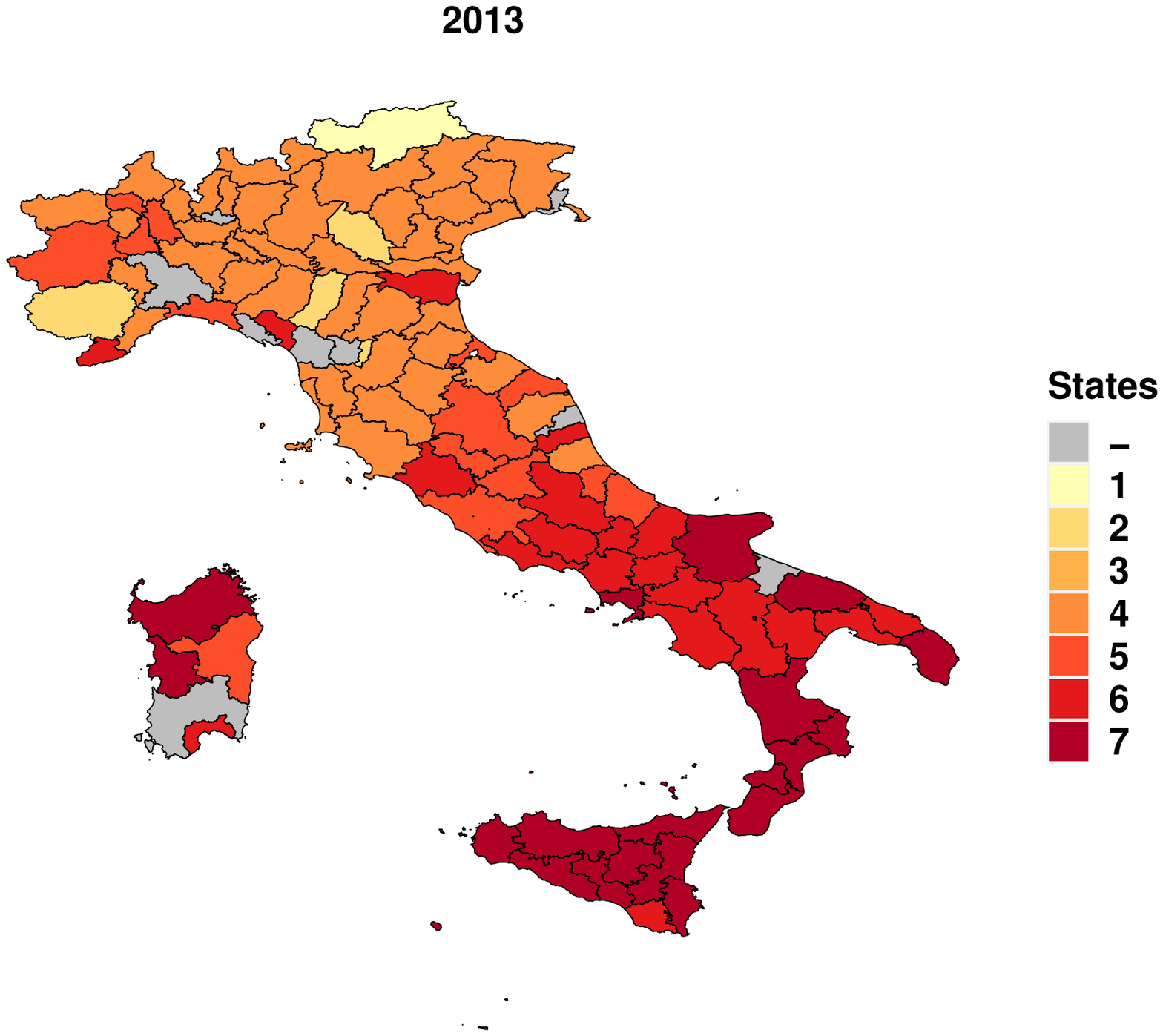}}
	\caption{Italian provinces map colored according to the estimated state memberships} 
\label{fig:6}       
\end{figure}
\begin{figure}[!htt]\ContinuedFloat
\subfloat[\label{fig:2018}]{
	\includegraphics[width=0.44\textwidth]{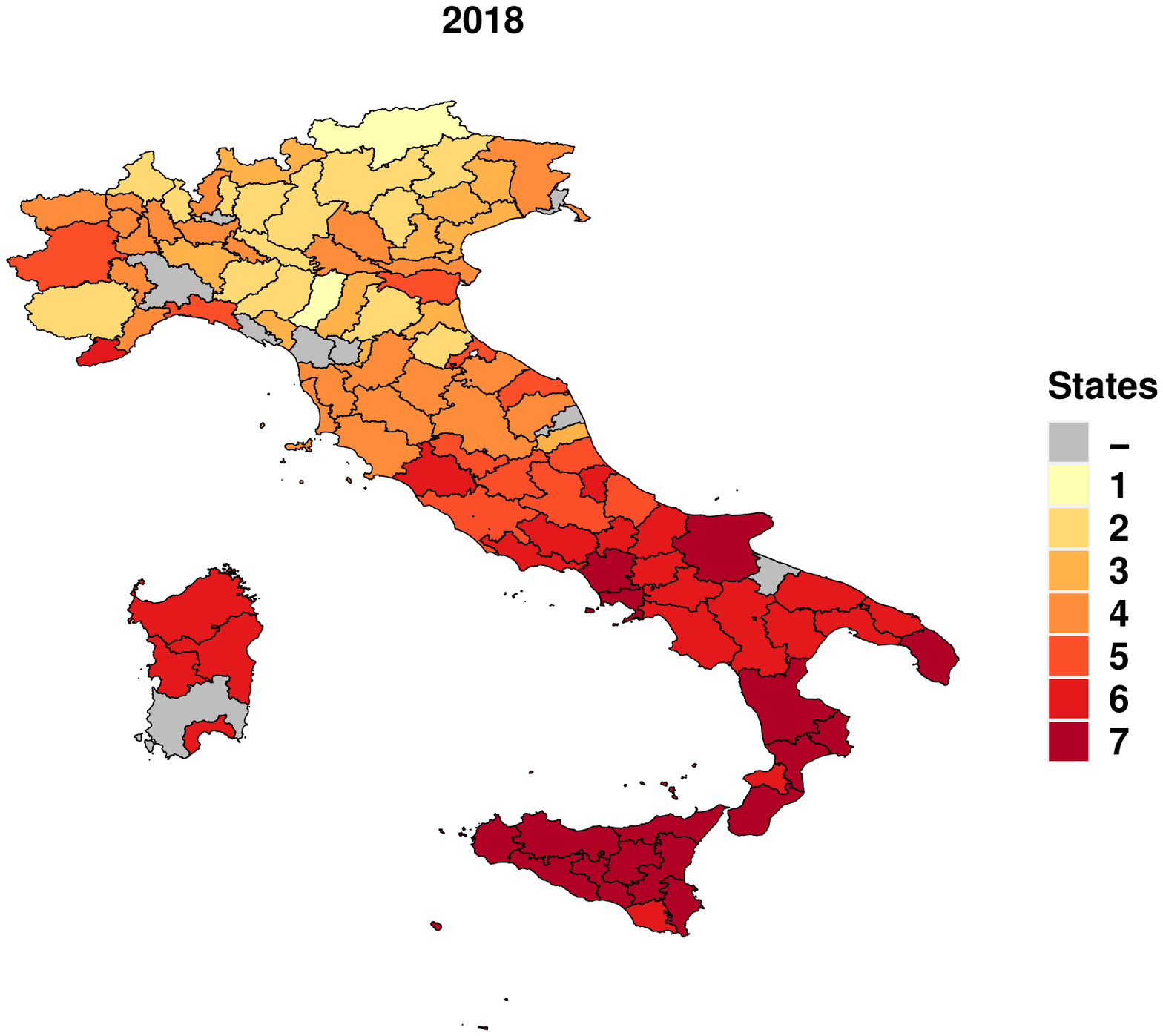}}\hspace{0.5em}
\subfloat[\label{fig:2019}]{%
	\includegraphics[width=0.44\textwidth]{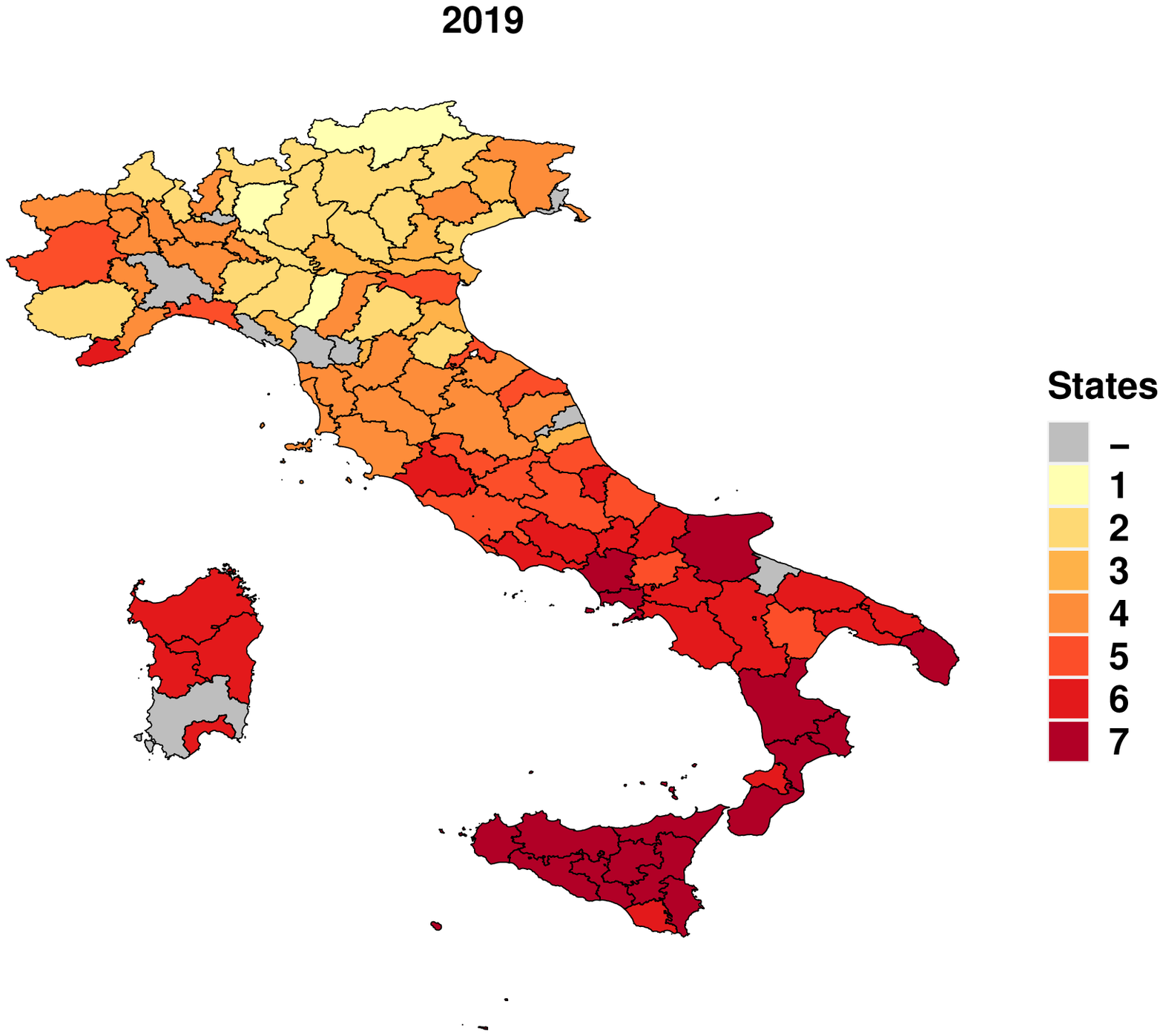}}
\caption{Italian provinces map colored according to the estimated state memberships (cont.)} 
\label{fig:7}       
\end{figure}

Starting from the first year of analysis, i.e.~2004, in Figure~\ref{fig:2004} we can recognize several clusters of provinces that, as we move towards the south, belong to states with higher unemployment rates.
After some years (2005--2008) characterized by relatively few changes among the states, the first economic recession began to produce its effects in 2009, where a lot of provinces started to perform badly (see Figure~\ref{fig:2009}).
Such event, further strengthened by the second economic recession of 2011, lead to a continuous rise of the unemployment levels, that (see \Cref{fig:2010,fig:2011,fig:2012,fig:2013}) lead the majority of provinces to the worst states, reducing the pre-existing differences between them.
After two years of small changes (2014--2015), the provinces located in the upper part of the Country started to recover, whereas the rest of the provinces failed to start again, remaining anchored in the post-recessions difficulties (see \Cref{fig:2018,fig:2019} that show the most two recent years).
In any case, these signs of recovery are going to be dramatically arrested by the COVID-19 pandemic, and its effects will have serious repercussions in the next years.

It is also interesting to report the behavior of some specific provinces.
For example, there is only one province that, over the 16 years, has never changed its state.
Specifically, the province of Bolzano, the northernmost in Italy, has been always in state one, the best.
This means that, regardless of the economic crises, the unemployment issue has been virtuously managed in this province.
We can also mention the two provinces that have worsened their position the most over the 16 years.
In detail, the province of Ancona and Ferrara were in state one in the first years of our analysis, but they are actually located in the fifth state (since 2013 and 2016, respectively).
Conversely, none of the provinces is actually in a state that can be considered better than the one had in 2004.
This means that all the provinces have kept (Bolzano), lost (most of the provinces) and at best regained (only some provinces) the state they had at the beginning of our analysis.

\section{Conclusions}
\label{sec:conc}

In this manuscript we introduced parsimonious hidden Markov models for matrix-variate longitudinal data.
Being (dependent) mixture models, they allow the recovery of homogenous latent subgroups and, simultaneously, provide meaningful interpretation on how the sample units move between the hidden states over time.
The parsimony has been introduced via the eigen decomposition of the state covariance matrices, producing a family of 98 HMMs.
An ECM algorithm has been illustrated for parameter estimation.
At first, the parameter recovery of our algorithm has been evaluated under different scenarios, providing good results.
This can be particularly interesting for those HMMs that use a MM algorithm at each step of the ECM algorithm.
Relatedly, we have analyzed the computational times for fitting all the 98 HMMs.
The computational burden of the HMMs using MM algorithm is definitely higher, even if we are able to fit all the HMMs in a quite fast way when parallel computing is considered.
The BIC has proven to be effective in detecting the true number of states in the data as well as the parsimonious structure.
The real data example has shown the usefulness of our HMMs.
Indeed, other than identifying different states, they have provided a tool for easily analyzing the evolution over time of the unemployment at province level, and for obtaining useful insight on the behavior of some specific provinces. 

There are different possibilities for further work, some of which are worth mentioning.
First of all, we can extend our HMMs by using skewed or heavy tailed state dependent probability density functions \citep{gallaugher2017matrix, gallaugher2019three, tomarchio2020two,tomarchio2020mixtures}, in order to model possible features commonly present in the data.
Another extension would deal with the regression setting \citep{viroli2012matrix}, where covariates shared by all units in the same hidden state are used.
This can be done both in a fixed and in random covariates framework.

%\begin{appendices}
%\section{}
%\label{sec:app}

%Let $\ddot \vecY=\sum_{k=1}^{K} \ddot \vecY_k$ be the update of the within cluster scattering matrix, where $\ddot \vecY_k=\sum_{i=1}^{N}\ddot z_{ik}\ddot w_{ik}\left(\bx_i-\ddot \matmu_k\right)\left(\bx_i-\ddot \matmu_k\right)^\top$ is the update of the scatter matrix associated with the $k$th component.
%The updates of $\vecSigma_k$ for the 14 parsimonious models are:

%\begin{itemize}
%\item Model EII [$\lambda\bI$]
%\begin{equation}
%\ddot\lambda = \frac{\tr\left\{\ddot \vecY\right\}}{Np}.
%\label{eq:eii}
%\end{equation}
%\end{itemize}
%\end{appendices}

\bibliographystyle{spbasic}      % basic style, author-year citations
\bibliography{HMM.bib}   % name your BibTeX data base

\end{document}